\documentclass[prl,twocolumn,preprintnumbers,superscriptaddress,amsmath,amssymb]{revtex4-1}

\usepackage[dvipdfmx]{graphicx}
\usepackage{epsfig}
\usepackage{subfigure}
\usepackage{mathrsfs}
\usepackage{amsfonts}
\usepackage{times}
\usepackage{amsmath}
\usepackage{leftidx}
\usepackage{tikz}
\usepackage{tikz-network}
\usepackage{color}
\usepackage[colorlinks,linkcolor=blue,citecolor=blue]{hyperref}

\newcommand{\Tr}{\operatorname{Tr}}

\usepackage{bbold}
\usepackage{braket}
\usepackage{mathtools}
\usepackage{colortbl}
\usepackage{multirow}

\begin{document}
\title{Any Quantum Many-Body State under Local Dissipation will be Disentangled in Finite Time}
\author{Zongping Gong}
\affiliation{Department of Applied Physics, University of Tokyo, 7-3-1 Hongo, Bunkyo-ku, Tokyo 113-8656, Japan}
\author{Yuto Ashida}
\affiliation{Department of Physics, University of Tokyo, 7-3-1 Hongo, Bunkyo-ku, Tokyo 113-0033, Japan}
\affiliation{Institute for Physics of Intelligence, University of Tokyo, 7-3-1 Hongo, Tokyo 113-0033, Japan}
\date{\today}

\begin{abstract}
%[YA: shall we be a bit more specific in title as above?]
%We show that under rather general assumptions entanglement sudden death occurs for arbitrary many-body state subject to local dissipation after an $\mathcal{O}(1)$ time threshold. 
%In many-body quantum spin systems subject to generic local dissipation, including depolarization, any initial state is found to become fully separable after a finite time independent of the system size. 
We prove that any quantum many-spin state under genetic local dissipation will be fully separable after a finite time independent of the system size. Such a sudden death of many-body entanglement occurs universally provided that there is a finite damping gap and the unique steady-state density matrix is of full rank. %We also construct a counterexample showing that the condition of strict locality cannot be loosen without further assumptions.  
This result is rigorously derived by combining a state-reconstruction identity based on random measurements and the convergence bound for quantum channels. Related works and possible generalizations are also discussed.

\end{abstract}
\maketitle

%\emph{Introduction.---}
Recognized as a fundamental non-classical feature of the nature \cite{RH09}, entanglement has been playing an increasingly important role in quantum many-body physics \cite{LA08,JE10,RI15,NL16,DAA19}. For isolated systems at zero temperature, %thermal equilibrium, 
a modern viewpoint on quantum phases of matter is that they are classified by qualitatively different entanglement patterns \cite{CKC16,XGW17,BZ19}. %For example, topological order can be defined as exhibiting long-range entanglement, meaning that the state cannot be disentangled by any finite-depth quantum circuit of local unitary gates. 
In nonequilibrium dynamics, the evolution of entanglement is known to be closely related to quantum thermalization and dynamical phase transitions \cite{RN15,AMK16,AL19,MPAF23}. From a practical perspective, the entanglement structure lies at the heart of numerical calculations based on tensor-network approaches, which have become prevalent in solving many-body problems in and out of equilibrium \cite{FV08,US11,JIC21}. 

\begin{figure*}
    \centering
    \includegraphics[width=0.75\textwidth]{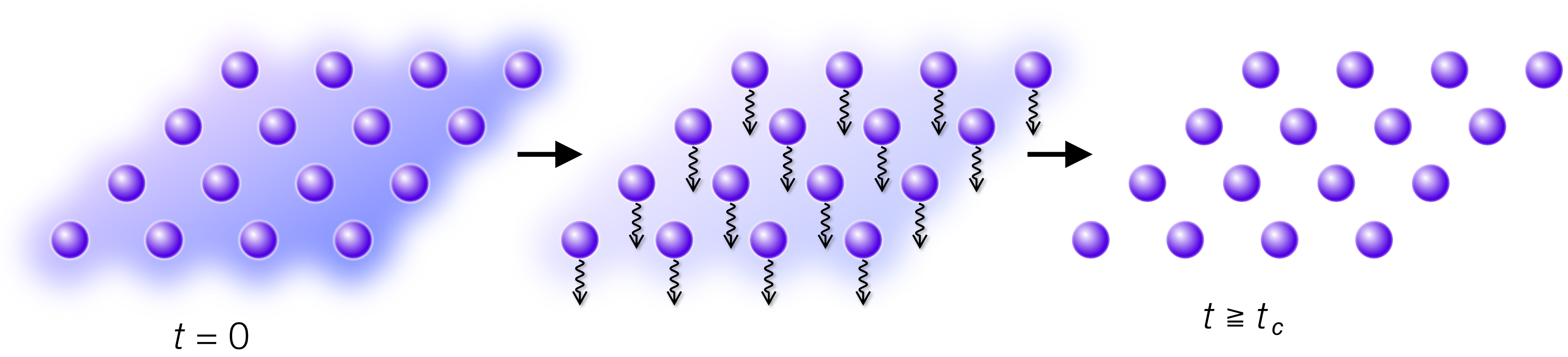}
    \caption{%Schematic illustration of entanglement sudden death 
    Schematic illustrating the sudden death of entanglement in quantum many-body systems. Due to local dissipation, an arbitrary initial state will be fully separable after a finite time $t_{\rm c}$  independent of the system size (cf.~Eq.~\eqref{tc}).}
    \label{fig:SD}
\end{figure*}

%Entanglement sudden death
%Recently, the paradigm of quantum many-body physics has greatly been extended to include open systems and mixed states. A prototypical setup in this context is to consider a many-body state subject to local dissipation and decoherence. %, such as isotropic depolarization accounting for white noise. 
%While the expectation value of an observable, a linear function of the density operator, should undergo exponential decay, 
Compared to pure states, it is much harder to analyze mixed-state entanglement in open quantum systems, which are ubiquitous in the real world \cite{LA15}. Even for two qubits subject to local dissipation that eventually leads to a separable steady state, the entanglement dynamics can be unexpectedly rich and counterintuitive \cite{TY04,TY06,BB07,LM10}. A remarkable observation is that entanglement may vanish at a finite time and stay exactly zero afterwards, a phenomenon known as the \emph{sudden death of entanglement} \cite{TY09}. This phenomenon was also observed for several specific many-body states and channels of particular interest in quantum information, such as the GHZ states, spin squeezed states, and cluster states under local decoherence or depolarization \cite{CS02,WD04,MH05,LD08,XW10}. 
Nevertheless, there remain fundamental questions: Does the entanglement sudden death  occur universally in a many-body system and, if so, how can one estimate the time threshold for arbitrary initial states under general dissipation?

%Motivation: generalize the origianl bipartite entanglement in few-level systems to multipartite entanglement in many-body systems. Mention the recent breakthrough by Ewin Tang, and highlight the inapplicability: not every mixed state is a Gibbs state; even yes, not necessarily local  
In this study, we show that, under minimal assumptions, %provided the steady state is full ranked, %and the thermodynamic limit is well-defined, 
a many-body state under local dissipation %evolved by local quantum channels 
will always become \emph{fully} separable after a \emph{finite} time %number of time steps 
independent of the system size. Here ``local'' is used in the quantum information sense to refer to operations performed on a single party \cite{VV97}. The threshold can be explicitly estimated from the relaxation properties of the quantum channels describing local dissipation. This result is valid regardless of the underlying %lattice structures, 
space dimension and configuration of partition. An initial state is also completely arbitrary and can be highly entangled. %such as obeying the volume law. 
The only requirement is that a local channel  has a unique, full ranked steady state. This is, however, %should be 
typically the case, as has been rigorously justified using random matrix theory \cite{LS20}. Hence, the full-rank assumption is no more than excluding some atypical exceptions. Indeed, it is known that sudden death may not occur in a system of two atoms undergoing spontaneous decay, leading to a pure steady state (thus not full ranked) \cite{TY04}.

\emph{Main result.---}We start from presenting a rigorous statement of our main result. Consider an $N$-party ($N\ge2$) quantum spin system with entire Hilbert space $\mathbb{H}=\bigotimes^N_{j=1}\mathbb{H}_j$. We denote $d_j=\dim \mathbb{H}_j$ ($j=1,2,...,N$) as the local Hilbert-space dimension of the $j$th party. %These dimensions may be different in general, as is especially the case when we group different numbers of qubits into different parties. 
We model local dissipation as a repetition of a nonunitary evolution generated by $\mathcal{E}=\bigotimes^N_{j=1}\mathcal{E}_j$, where $\mathcal{E}_j$ is a quantum channel acting locally on the $j$th party. Each channel $\mathcal{E}_j$ is assumed to have a unique fixed point $\hat\pi_j=\mathcal{E}_j(\hat\pi_j)$ with full rank. We denote $\mu_j$ as the second largest normed eigenvalue of $\mathcal{E}_j$, and $\lambda_j$ as the smallest eigenvalue of $\hat{\pi}_j$, respectively. Throughout this Letter, we express discrete time $t\in\mathbb{N}$ in the unit of the time period of each channel $\cal E$. Note that stroboscopic dynamics generated by local Lindbladians are included as a special case, in which $\hat{\pi}_j$ and $-\ln\mu_j$ are nothing but the steady state and the Liouvillian gap, respectively. %\textcolor{red}{Here the period of stroboscope is set to be unit.} 
Here we consider a discrete-time setup to cover all possible local channels that do not necessarily admit Lindblad generators \cite{JIC08}. %expressions.

Our main result is the following: For \emph{any} initial state $\hat\rho(0)=\hat\rho$ on $\mathbb{H}$, $\hat\rho(t)=\mathcal{E}[\hat\rho(t-1)]=...=\mathcal{E}^{t-1}(\hat\rho(1))=\mathcal{E}^t(\hat\rho)$ becomes fully separable, i.e., takes the form of \cite{MBP07,MW16,PH24}
\begin{equation}
\sum_{\boldsymbol{\alpha}=\alpha_1\alpha_2...\alpha_N}P_{\boldsymbol{\alpha}}
\hat\rho^{(\alpha_1)}_1\otimes\hat\rho^{(\alpha_2)}_2\otimes...\otimes\hat\rho^{(\alpha_N)}_N
%\sum_{\boldsymbol{\alpha}} P_{\boldsymbol{\alpha}} \bigotimes^N_{j=1}\hat\rho^{(\alpha_j)}_j,
\label{fs}
\end{equation}
for some density operator $\hat\rho^{(\alpha)}_j$ on $\mathbb{H}_j$ and $P_{\boldsymbol{\alpha}}\ge0$ ($\sum_{\boldsymbol{\alpha}}P_{\boldsymbol{\alpha}}=1$), when the dimensionless time satisfies $t\ge \lceil t_{\rm c}\rceil$ with
\begin{equation}
t_{\rm c}=\max_j t_j,\;\;\;\;
t_j= \kappa_j^{-1}\ln[C_j\lambda_j^{-2}(d_j+\lambda_j)].
\label{tc}
\end{equation}
See Fig.~\ref{fig:SD} for a schematic illustration. Here in Eq.~(\ref{tc}), positive constants $C_j$ and $\kappa_j$ are identified from the convergence bound for the $j$th local channel:
\begin{equation}
\|\mathcal{E}^t_j(\hat\rho_j) - \hat\pi_j\|%_1
\le C_j e^{-\kappa_j t},
\label{cb}
\end{equation}
where $\hat\rho_j$ is an arbitrary initial state on $\mathbb{H}_j$ and $\|\cdot\|$ is the operator norm given by the maximal singular value \footnote{One can adopt other norms, such as the trace norm more commonly used in the literature \cite{MJK12,OS15}. However, since we focus on a \emph{local} Hilbert space of finite dimension, different norms only differ by a constant factor independent of $N$ \cite{RAH12}, thus the convergence bound (\ref{cb}) will have no formal difference.}. %$\|\hat \sigma\|_1=\Tr[\sqrt{\hat \sigma^\dag \hat \sigma}]$ is the trace norm. 

We emphasize that the Hilbert-space dimensions of individual parties may not be the same, let alone the local channels. %In this case, 
To make our statement compatible with the thermodynamic limit, we require the parameters $d_j$, $\mu_j$, $\lambda_j$ to be uniformly bounded. That is, denoting $d=\max^N_{j=1}d_j$, $\mu=\max^N_{j=1}\mu_j$, $\lambda=\min^N_{j=1}\lambda_j$, we always have $d<\infty$, $\mu<1$ and $\lambda>0$ even %after taking 
in the limit of $N\to\infty$. In particular, $\mu<1$ means that the global channel $\mathcal{E}$ has a finite damping gap. %Clearly, one need not care about this subtlety in the homogeneous case with no $j$ dependence.  %where everything does not depend on  

A few comments are in order. The threshold in Eq.~(\ref{tc}) is just an upper bound on the actual sudden-death time. In other words, the state may already become fully separable before $t_{\rm c}$. A trivial example is when the state is fully separable from the very beginning. The nontrivial prediction is that, even if $\hat\rho$ is highly entangled like exhibiting long-range \cite{AK06,ML06,HCJ12} or even voulme-law entanglement \cite{DNP93,SP06,EB22}, the sudden death still occurs at a finite time independent of the system size. We would also like to mention that Eq.~(\ref{cb}) is not so obvious as it appears. In fact, it cannot be derived from the usual spectral decomposition, and great efforts are required to tighten %it could be very technical to optimize 
$C_j$ \cite{MJK12,OS15}. We will soon encounter a similar difficulty in analyzing the disentangler structure (cf. Eq.~(\ref{Et})) of $\mathcal{E}_j^t$. In the worst case, $C_j$ may blow up with increasing $d_j$ %and the worst scaling is 
as $e^{\mathcal{O}(d_j^2\ln d_j)}$. Also, while one may na\"ively take $\kappa_j=-\ln \mu_j$, Eq.~(\ref{cb}) may no longer be valid for such a choice if $\mathcal{E}_j$ is not diagonalizable. One should then choose $\kappa_j$ to be strictly smaller than $-\ln \mu_j$ \cite{MF92}. %Nevertheless, the point is that $C_j$ is finite and $\kappa_j$ is strictly positive and both of them can be determined solely from $d_j$ and $\mu_j$.  %will be finite provided $d<\infty$ and $\mu<1$. 
Nevertheless, the important point is that, suppose the parameters are uniformly bounded, the threshold (\ref{tc}) is finite even in the thermodynamic limit.
%This fact is actually not so trivial as it appears. In particular

\emph{Case study of depolarization channel.---}It is instructive to briefly discuss the case of qubits under local depolarization, a widely studied example in the literature \cite{CS02,WD04,MH05,LD08,XW10}. In this case, all the local channels are chosen to be
\begin{equation}
\mathcal{E}_j(\hat\rho_j)=p\hat\rho_j+\frac{1}{2}(1-p)\hat\sigma_0\Tr[\hat\rho_j],\;\;\forall j=1,2,...,N,
\label{dp}
\end{equation}
where parameter $p\in(0,1)$ measures the depolarization strength (stronger for smaller $p$) and $\hat\sigma_0$ is the qubit identity. %As noted in Ref.~\cite{CS02}, 
The channel with $p=1/3$ can be realized by random projective measurement \cite{CS02}, as can be confirmed using the Weingarten calculus \cite{DW78,PWB96,BC06}. Since $\mathcal{E}_j^t$ takes the same form (\ref{dp}) but with $p$ replaced by $p^t$, %Applying this result to the global channel $\mathcal{E}$, 
we know that any input will become fully separable above $\lceil t_{\rm c}\rceil=\lceil - \ln 3/\ln p\rceil$. This is because in the random measurement unraveling of Eq.~(\ref{dp}) with $p=1/3$, each measurement outcome corresponds to a (pure) product state, and the ensemble average turns out to be a statistical mixture of these product states. This analysis can be readily generalized to qudit systems \cite{MH99} and the corresponding time threshold reads %$p=1/(d+1)$  
$\lceil t_{\rm c}\rceil=\lceil -\ln(d+1)/\ln p\rceil$. We mention that the threshold given by Eq.~(\ref{tc}) is slightly looser than this value \footnote{For the qudit depolarization channel, we have $\|\mathcal{E}^t(\hat\rho)-\hat{\mathbb{1}}/d\|= p^t\|\hat\rho-\hat{\mathbb{1}}/d\|\le (d-1)p^t/d$, so we can take $C=(d-1)/d$ and $\kappa=-\ln p$. Combining with $\lambda=1/d$, we obtain $t_{\rm c}=-\ln[(d-1)(d^2+1)]/\ln p$.}, consistent with the fact that Eq.~(\ref{tc}) always upper bounds the actual sudden death time.

\emph{Informal %statement
argument.---}We start from giving an informal argument that indicates the entanglement sudden death for a general local quantum channel. Clearly, we have a fully separable output when performing arbitrary local projective measurements, which are not necessarily Haar random and may be followed by feedback control. The latter does not create entanglement if executed locally \cite{VV97}. In this case, a local channel can be expressed as $\mathcal{E}_j(\hat\rho_j)= \sum_\alpha p_\alpha \hat U_\alpha\hat \Pi_\alpha \hat\rho_j \hat\Pi_\alpha\hat U_\alpha^\dag$ 
%\hat U^{(\alpha)}_j\hat\Pi^{(\alpha)}_j \hat\rho_j \hat\Pi^{(\alpha)}_j \hat U^{(\alpha)\dag}_j$
%\label{UPi}
with $\hat\Pi_\alpha$ %$\hat \Pi^{(\alpha)}_j$ 
a rank-$1$ projector satisfying $\sum_\alpha p_\alpha\hat\Pi_\alpha %\hat\Pi^{(\alpha)}_j
=\hat{\mathbb{1}}_j$ ($p_\alpha\ge0$), 
and $\hat U_\alpha$ %$\hat U^{(\alpha)}_j$ 
a unitary feedback protocol conditioned on outcome $\alpha$ \cite{TS08}. It is helpful to rewrite the expression of $\mathcal{E}_j$ into
\begin{equation}
\mathcal{E}_j(\hat\rho_j)=\sum_\alpha\hat\rho^{(\alpha)}_j\Tr[\hat E^{(\alpha)}_j\hat\rho_j],
\label{rE}
\end{equation}
where $\hat\rho^{(\alpha)}_j=\hat U_\alpha \hat\Pi_\alpha \hat U_\alpha^\dag$ %\hat U^{(\alpha)}_j\hat\Pi^{(\alpha)}_j\hat U^{(\alpha)\dag}_j$ 
and $\hat E^{(\alpha)}_j= p_\alpha\hat \Pi_\alpha$. %p_\alpha\hat\Pi^{(\alpha)}_j$. 
In fact, Eq.~(\ref{rE}) can be made more general, in the sense that $\hat\rho^{(\alpha)}_j$ can be any density operator and $\{\hat E^{(\alpha)}_j\}_\alpha$ can be any positive operator-valued measure \cite{MAN10} on the $j$th party. %Once we can show 
If each local channel takes the form of Eq.~(\ref{rE}), the output is clearly fully separable -- it takes the form of Eq.~(\ref{fs}) with $P_{\boldsymbol{\alpha}}=\Tr[\hat E^{(\boldsymbol{\alpha})}\hat\rho]$, where $\hat E^{(\boldsymbol{\alpha})}=\hat E^{(\alpha_1)}_1\otimes\hat E^{(\alpha_2)}_2\otimes...\otimes \hat E^{(\alpha_N)}_N$.  %$\mathcal{E}(\hat\rho)=\sum_{\boldsymbol{\alpha}} p_{\boldsymbol{\alpha}}\hat\rho_{\boldsymbol{\alpha}}$. 
%Here $\boldsymbol{\alpha}$ is a length-$N$ string with the $j$th letter taking on $\alpha$ in Eq.~(\ref{rE}), $p_{\boldsymbol{\alpha}}=\Tr[\hat E_{\boldsymbol{\alpha}}\hat\rho]$, and $\hat\rho_{\boldsymbol{\alpha}}$ ($\hat E_{\boldsymbol{\alpha}}$) is a direct product of $\hat\rho^{(\alpha)}_j$ ($\hat E^{(\alpha)}_j$).

%Now we only have 
It thus suffices to show that $\mathcal{E}^t_j$ always takes the form of Eq.~(\ref{rE}) for a sufficiently large $t$. %Obviously $\mathcal{E}^\infty_j(\;\cdot\;)=\hat\pi_j\Tr[\;\cdot\;]$ does take this form. 
For now, %the sake of simplicity, 
we assume $\mathcal{E}_j$ to be diagonalizable so that
%\begin{equation}
$\mathcal{E}^t_j(\hat\rho_j)=\hat\pi_j\Tr[\hat\rho_j]+\sum_{\alpha} \xi_\alpha^t \hat r_\alpha\Tr[\hat l_\alpha^\dag\hat\rho_j]$, 
%\label{Et}
%\end{equation}
where $\hat r_\alpha$ and $\hat l_\alpha$ are a pair of left and right eigen-operators of $\mathcal{E}_j$ %(super)vectors 
with eigenvalue $\xi_\alpha$, and $\mu_j=\max_\alpha|\xi_\alpha|<1$ \cite{KM16}. Since the channel is Hermiticity-preserving, we have another pair of eigen-operators %(super)vectors 
$\hat r_{\alpha'}\propto\hat r_\alpha^\dag$ and $\hat l_{\alpha'}\propto\hat l_\alpha^\dag$ with eigenvalue $\xi_{\alpha'}=\xi_\alpha^*$ whenever $\alpha'\neq\alpha$. Here we write $\propto$ because of the guage invariance $\hat r_\alpha\to c\hat r_\alpha$, $\hat l_\alpha^\dag\to c^{-1}\hat l_\alpha^\dag$ $\forall c\in\mathbb{C}\backslash\{0\}$. Introducing $\hat R_\alpha(t)=(\xi^t_\alpha\hat r_\alpha + \xi^{*t}_\alpha\hat r_\alpha^\dag)/(\sqrt{2}|\xi_\alpha|^t)$, $\hat R_{\alpha'}(t)=(\xi^t_\alpha\hat r_\alpha - \xi^{*t}_\alpha\hat r_\alpha^\dag)/(\sqrt{2}i|\xi_\alpha|^t)$, $\hat L_\alpha=(\hat l_\alpha^\dag + \hat l_\alpha)/\sqrt{2}$ and $\hat L_{\alpha'}=(\hat l_\alpha^\dag - \hat l_\alpha)/(\sqrt{2}i)$ for those $\alpha'\neq\alpha$ (otherwise $\hat R_\alpha=\hat R^\dag_\alpha\propto\hat r_\alpha$, $\hat L_\alpha=\hat L^\dag_\alpha\propto\hat l_\alpha$), which are all Hermitian (and $\hat R_\alpha$ being traceless), we can recast the spectral decomposition of $\mathcal{E}^t_j$ into 
\begin{equation}
\begin{split}
\mathcal{E}^t_j(\hat\rho_j)&=\hat\pi_j\Tr[\hat\rho_j]+\sum_{\alpha} |\xi_\alpha|^t \hat R_\alpha(t)\Tr[\hat L_\alpha \hat\rho_j],\\
&=\sum_{\alpha,s=\pm}\hat \rho_{\alpha s}(t)\Tr[\hat E_{\alpha s}\hat\rho_j],
\end{split}
\label{Et}
\end{equation}
where $\hat E_{\alpha\pm}=(\hat{\mathbb{1}}_j\pm \hat L_\alpha)/(2D_j)$ with $D_j= d_j^2-1$, and 
\begin{equation}
\hat \rho_{\alpha\pm}(t)=\hat\pi_j\pm D_j|\xi_\alpha|^t\hat R_\alpha(t).
\label{rat}
\end{equation}
%being the total number of $\alpha$. 
%, i.e., that %the number of nonzero eigenvalues of $\mathcal{E}_j$ not equal to $1$. 
%Since there is a gauge invariance $\hat r_\alpha\to c\hat r_\alpha$, $\hat l_\alpha^\dag\to c^{-1}\hat l_\alpha^\dag$ $\forall c\in\mathbb{C}\backslash\{0\}$ 
Using the gauge redundancy in the spectral decomposition, 
we can fix the norm of $\hat l_\alpha$ such that $\|\hat L_\alpha\|\le1$ %($\|\cdot\|$: operator norm) 
and $\hat E_{\alpha\pm}\ge0$. Recalling the full-rank assumption for $\hat\pi_j$, we know that $\hat\rho_{\alpha\pm}(t)$ in Eq.~(\ref{rat}) will always become a density operator for a sufficiently large $t$, since %the factor 
$|\xi_\alpha|^t$ converges to zero as  $t$ is increased. %can be made arbitrarily small. 

Unfortunately, the above argument is not quite general as it %concerning the structure of late-time channels 
relies on the diagonalizability, i.e., away from an exceptional point \cite{WDH12}. Moreover, even if $\mathcal{E}_j$ is diagonalizable but extremely close to an exceptional point, $\|\hat r_\alpha\|$ and $\|\hat R_\alpha(t)\|$ may blow up to an arbitrarily large value \cite{YA21} for given $d_j$ and $\mu_j$, leading to an unbounded threshold. As mentioned previously, the same difficulty appears if one tries to derive a convergence bound (\ref{cb}) using the spectral decomposition \cite{MJK12,OS15}. 
%for entanglement sudden death even for fixed $d$ and $\mu$. 
In the following, we will provide a general %an alternative 
proof that focuses more on the state $\hat\rho(t)$ itself. %rather than the channel. 
We can then bypass the difficulty %from nondiagonalizability 
by leveraging the well-established convergence bound (\ref{cb}). %which has been solved in previous studies \cite{MJK12,OS15}.

\emph{Proof.---}Before delving into %introducing 
the %most 
general proof, we first explain how to reproduce the $p=1/3$ threshold for qubit systems under depolarization (\ref{dp}). As mentioned above, we now focus more directly on the state. If we are interested in the entanglement between the $j$th qubit and the remaining, we can perform a general decomposition
%\begin{equation}
$\hat\rho= \sum_{\mu=0,x,y,z}\hat\sigma_\mu \otimes \hat\Sigma_\mu$,
%\end{equation}
where the Pauli operator $\hat\sigma_\mu$ acts on the $j$th qubit and $\hat\Sigma_\mu=\hat\Sigma_\mu^\dag$ acts on the remaining $N-1$ qubits. We can recast this decomposition into
\begin{equation}
\begin{split}
\hat\rho&= \frac{1}{6}\sum_{\mu=x,y,z,s=\pm}(\hat\sigma_0+3s\hat\sigma_\mu)\otimes(\hat\Sigma_0+s\hat\Sigma_\mu)\\
&=\frac{1}{3}\sum_{\mu=x,y,z,s=\pm}(3\hat\Pi_{\mu s}-\hat\sigma_0)\otimes\Tr_j[\hat\Pi_{\mu s}\hat\rho],
\end{split}
\label{qb}
\end{equation}
where $\hat\Pi_{\mu\pm}=(\hat\sigma_0\pm\hat\sigma_\mu)/2$ is a projector %$\hat\Sigma_0\pm\hat\Sigma_\mu =\Tr_j[\hat\rho(\hat\sigma_0\pm\hat\sigma_\mu)]/2$ turns out to be always positive semi-definite 
and thus $\hat\Sigma_0\pm\hat\Sigma_\mu=\Tr_j[\hat\Pi_{\mu\pm}\hat\rho]$ is a density operator up to normalization. Applying locally $\mathcal{E}_j$ in Eq.~(\ref{dp}) to $\hat\rho$, we find $3\hat\Pi_{\mu s} - \hat\sigma_0$ %$(\hat\sigma_0\pm3\hat\sigma_\mu)/2$ 
in Eq.~(\ref{qb}) is turned into a density operator $3p\hat\Pi_{\mu s}+(1-3p)\hat\sigma_0/2$ %$(\hat\sigma_0\pm3p\hat\sigma_\mu)/2\ge0$ 
for $p\le1/3$, implying that the $j$th party is disentangled from the remaining. Repeating the above procedures for the remaining parties, we will obtain %finally end up with 
a fully separable expression for $\mathcal{E}(\hat\rho)$.

%\textcolor{red}{key point: unitary 2-design}

To generalize the above analysis, we have to figure out the qudit counterpart of Eq.~(\ref{qb}). %, say $\hat\rho= \sum_\alpha \hat m_\alpha \otimes \hat M_\alpha$, where $\hat m_\alpha$ ($\hat M_\alpha$) is Hermitian and acts on the $j$th party (remaining parts). Here we require $\Tr\hat m_\alpha>0$ and $\hat M_\alpha\ge0$.
This turns out to be realizable by an arbitrary \emph{unitary $2$-design} $\{\hat U^{(\alpha)}_j\}_\alpha$ for a single qudit \cite{DG07}. Given any reference state $|0\rangle$ in $\mathbb{H}_j$, we define a set of rank-$1$ projectors $\hat\Pi^{(\alpha)}_j=\hat U^{(\alpha)}_j |0\rangle\langle0|\hat U^{(\alpha)\dag}_j$ in terms of the unitary $2$-design. %This set of projectors suffices 
They suffice to reproduce the effect of %completely 
Haar random projective measurement on the ensemble level:
\begin{equation}
\frac{d_j}{D_j}\sum_\alpha \hat\Pi^{(\alpha)}_j\hat\rho\hat\Pi^{(\alpha)}_j
= \frac{1}{d_j+1}(\hat\rho + %\frac{1}{d_j + 1}
\hat{\mathbb{1}}_j\otimes\Tr_j[\hat\rho]),
\end{equation}
where $D_j$ is the cardinality of $\{\hat U^{(\alpha)}_j\}_\alpha$. Since a $2$-design is always %automatically 
a $1$-desgin, we have $D_j^{-1}\sum_\alpha\hat\Pi^{(\alpha)}_j=d_j^{-1}\hat{\mathbb{1}}_j$ and 
\begin{equation}
\hat \rho = \frac{d_j}{D_j}\sum_\alpha [(d_j+1)\hat\Pi^{(\alpha)}_j-\hat{\mathbb{1}}_j]\otimes\Tr_j[\hat\Pi^{(\alpha)}_j\hat\rho].
\end{equation}
Note that Eq.~(\ref{qb}) is reproduced in the qubit case by taking the $2$-design to be the single-qubit Clifford gates \cite{CD09} and $|0\rangle$ to be an eigenstate of $\hat\sigma_\mu$ ($\mu=x,y,z$). Alternatively, one may also directly require $\{\hat\Pi^{(\alpha)}_j\}_\alpha$ to form a spherical $2$-design \cite{JMR04}. We can repeat such a procedure of state reconstruction for the remaining parties and end up with
\begin{equation}
\hat\rho=\sum_{\boldsymbol{\alpha}} P_{\boldsymbol{\alpha}}\bigotimes^N_{j=1}[(d_j+1)\hat\Pi^{(\alpha_j)}_j-\hat{\mathbb{1}}_j],
\label{rP}
\end{equation}
where $P_{\boldsymbol{\alpha}}=(\prod^N_{j=1}d_jD_j^{-1})\Tr[\hat\Pi^{(\boldsymbol{\alpha})}\hat\rho]\ge0$ with $\hat\Pi^{(\boldsymbol{\alpha})}=\hat\Pi^{\alpha_1}_1\otimes\hat\Pi^{\alpha_2}_2\otimes...\otimes\hat\Pi^{\alpha_N}_N$. Note that identities similar to Eq.~(\ref{rP}) %similar decompositions 
have appeared in the context of shadow tomography \cite{HY20,SC21,VV24,BV24}, where random measurements also play a central role. If we again consider local depolarization, %then 
the threshold reads $p=\min_j 1/(d_j+1)$ since $(d_j+1)\hat\Pi^{(\alpha)}_j - \hat{\mathbb{1}}_j$ is turned into $(d_j+1)p\hat\Pi^{(\alpha)}_j + [1-(d_j+1)p]\hat{\mathbb{1}}_j/d_j$.

The remaining issue is to analyze general %generalize the dipolarization channel to an arbitrary 
channels with full-rank steady states. This can be %readily 
done with the help of the convergence bound (\ref{cb}). For an arbitrary Hermitian operator $\hat h$ on $\mathbb{H}_j$ with unit trace, %$w=\Tr\hat h>0$ and norm $g=\|\hat h\|$, 
we %can 
construct a density operator $\hat\rho_h=(\|\hat h\|\hat\pi_j+\lambda_j \hat h)/(\|\hat h\|+\lambda_j)$, whose positive semi-definiteness follows 
%from Weyl's perturbation theorem \cite{RB97}. 
from $\hat\pi_j\ge\lambda_j\hat{\mathbb{1}}_j$ and $\hat h\ge -\|\hat h\|\hat{\mathbb{1}}_j$.  
Substituting $\hat\rho_h$ into Eq.~(\ref{cb}) yields the inequality
\begin{equation}
\|\mathcal{E}_j^t(\hat h) - \hat\pi_j\|\le(1+\lambda_j^{-1}\|\hat h\|)C_je^{-\kappa_j t},
\end{equation}
which ensures the positive semi-definiteness of $\mathcal{E}_j^t(\hat h)$ for any $t\ge \lceil \kappa_j^{-1}\ln[C_j\lambda_j^{-2}(\|\hat h\|+\lambda_j)]\rceil$ as follows:
\begin{equation}
\begin{split}
\langle\psi|\mathcal{E}_j^t(\hat h)|\psi\rangle&\ge \langle\psi|\hat\pi_j|\psi\rangle -\|\hat\pi_j-\mathcal{E}_j^t(\hat h)\|\\
&\ge \lambda_j - (1+\lambda_j^{-1}\|\hat h\|)C_j e^{-\kappa_j t}\ge0,
\end{split}\label{semipos}
\end{equation}
where $|\psi\rangle$ is an arbitrary state in $\mathbb{H}_j$.
We now consider a choice $\hat h=\hat h^{(\alpha)}_j=(d_j+1)\hat\Pi^{(\alpha)}_j-\hat{\mathbb{1}}_j$ %we have $w=1/(d_j+1)$ and $g=d_j/(d_j+1)$ for $\hat h=\hat\Pi^{(\alpha)}_j - \mathbb{1}_j/(d_j+1)$, so that $w^{-1}g=d_j$ and 
with $\|\hat h\|=d_j$, leading to the threshold  $t_j=\lceil \kappa_j^{-1}\ln[C_j\lambda_j^{-2}(d_j+\lambda_j)]\rceil$. %Recalling that $d_j$ ($\lambda_j$) is uniformly upper (lower) bounded by $d$ ($\lambda$), we finally obtain the global threshold given in Eq.~(\ref{tc}).
From Eqs.~(\ref{rP}) and \eqref{semipos}, we have shown that $\mathcal{E}^t(\hat\rho)=\sum_{\boldsymbol{\alpha}}P_{\boldsymbol{\alpha}}\bigotimes^N_{j=1}\mathcal{E}_j^t(\hat h^{(\alpha_j)}_j)$ is fully separable when $t\ge t_{\rm c}=\max_j t_j$. This completes the proof of the main result.

\emph{Discussions.---}We emphasize that our essential progress lies in the size (i.e., %or equivalently, 
$N$) independence of $t_{\rm c}$,  
which remains finite in the thermodynamic limit $N\to\infty$. In fact, in any finite-level quantum system, the existence of the entanglement sudden death is ensured by the geometric structure of entanglement \cite{KZ98,GV99,SLB99}. That is, near any fully separable density operator with full rank, there is a finite-radius \emph{separable ball} in the operator space within which any operator corresponds to a fully separable state. In particular, choosing this separable state to be the steady state of $\mathcal{E}$, we know that $\mathcal{E}^t(\rho)$ will enter this separable ball for a sufficiently large $t$. However, the radius is exponentially small in $N$ \cite{RYW23}, so the time threshold will be proportional to $N$ in an exponential relaxation. Our result implies that sudden death occurs %the state already becomes separable 
somewhere far away from the separable ball of the steady state.     

This geometric picture applies equally to ``imaginary-time'' dynamics, i.e., %a family of 
Gibbs states with different temperatures \cite{GP24}. The infinite-temperature state is obviously fully separable and full-ranked, thus has a separable ball for a finite $N$, implying a finite-temperature sudden death \cite{BVF05}. Again, this threshold based on the separable ball will have an unsatisfactory scaling in $N$, while a size-independent threshold has recently been proved for general short-range interacting qubits \cite{AB24}. 
Nevertheless, we emphasize that our result cannot be implied by Ref.~\cite{AB24} simply because our initial state is arbitrary and thus a non-Gibbs state in general. For instance, it can even stay volume-law correlated (measured by mutual information) at any finite time, as is the case for a rainbow state \cite{GR15} under local depolarization. These states cannot be described as Gibbs states of short-range interacting Hamiltonians, which should always be area-law correlated \cite{MMW08}. 
We also mention an apparent contradiction with an example in Ref.~\cite{YK24} showing no sudden death. The point is that Ref.~\cite{YK24} has considered the projection onto a given parity sector of the Ising Hamiltonian, resulting in states that are not full-ranked. 
%(always full-ranked). %In particular, even the ``infinite-temperature'' state  is located at the boundary of the separable region.   %\textcolor{red}{[Apparent counterexample by Vijay]}
%We note that at the temperature threshold given in Ref.~\cite{AB24}, the Gibbs state is always short-range correlated \cite{MK14}.  In contrast, in our real-time setting, it may happen that the state remains long-range correlated despite being fully separable, as exemplified by the GHZ state under local depolarization. This example clearly shows that our result cannot be a corollary of Ref.~\cite{AB24}. 

\begin{figure}
    \centering
    \includegraphics[width=0.4\textwidth]{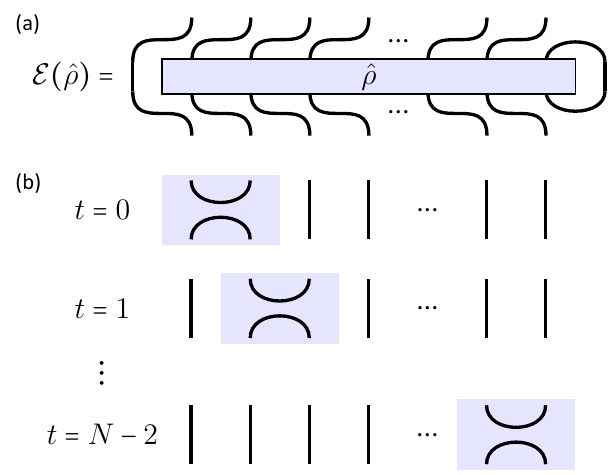}
    \caption{(a) Graphic representation of a one-dimensional %locality-preserving 
    channel with maximal damping gap and a unique (infinite-temperature) steady state of full rank. This channel preserves short-range correlation. (b) Persistence of two-qudit entanglement (shaded) up to $t=N-2$ for a specific input. In both (a) and (b), normalization constants are dropped for simplicity.}
    \label{fig:CE}
\end{figure}

Another natural question is whether or not our result can be generalized to not-strictly-local dissipation %quantum channel 
that may correlate adjacent parties, yet the steady state is fully separable or even a direct product. We claim that without %adding 
further conditions, our result is no longer valid in general. A simple counterexample is a one-dimensional translation combined with a full depolarization at a certain site on a ring of qudits, as shown in Fig.~\ref{fig:CE}(a). If we input a two-qudit entangled state far away from the depolarization site tensored with any remaining state %such as 
(e.g., the infinite-temperature state), the entanglement persists up to $\mathcal{O}(N)$ time (see Fig.~\ref{fig:CE}(b)), so the sudden-death threshold diverges in the thermodynamic limit. Note that the channel preserves short-range correlation \cite{LP20}, and actually maximally gapped, in the sense that its spectrum consists of a single $1$ and otherwise $0$. The latter can be seen from the fact that any generalized Pauli string, which constitues a complete orthonormal many-body operator basis, is irreversibly shifted to another string by the action of $\mathcal{E}$, with the only exception being the global identity.  It remains an important open problem what is the minimal additional condition that allows us to go  beyond local dissipation. Indeed, the condition for a fully separable steady state  already seems to be highly nontrivial. %Rapid mixing \cite{TSC15} might be a candidate as it rules out the counterexample mentioned above.

%Other non-classical measure, discord

\emph{Summary and outlook.---}In summary, we have demonstrated that the sudden death of entanglement occurs universally for any many-body states under generic local dissipation. By generic we mean a finite damping gap and the unique steady state is of full rank. The intuition on the channel side is that long-time evolution takes the form of a disentangler (\ref{rE}). The intuition on the state side is that we can always decompose it into a statistical mixture of tensored unit trace operators (\ref{rP}). The latter will be turned into a density operator %(up to normalization) 
by a generic channel after a size-independent finite %number of 
time. %steps. 

As discussed above, an urgent problem is to consider more general quantum channels beyond local dissipation. A good starting point might be combing local depolarization with Hamiltonian or quantum circuit evolution, which does not change the steady state (infinite temperature state) \cite{HP13,BS19,YF20}. Another direction is to consider quantum information-theoretic quantities other than entanglement, such as magic \cite{VV14,MH14,ZWL22}, discord \cite{HO01,AD08,CR20}, and Bell non-locality \cite{NB14,JT14,RS16}. One may also study the refined structures of entanglement, such as when %does 
long-range entanglement becomes short-ranged, signaling the death of topological order \cite{YHC24}. 
%The results may change a lot as some quantities may be nonzero even for fully separable states, and some may be zero even despite nonzero entanglement (e.g., Bell non-locality).  
%which is of relevance to many NISQ experiments.

We are grateful to Masahiro Hoshino, Naoki Iyama, Xiao-Qi Sun, and Shun Umekawa for fruitful discussions. 
Z.G. acknowledges support from the University of Tokyo Excellent Young Researcher Program and from JST ERATO Grant Number JPMJER2302, Japan. Y.A. acknowledges support from the Japan Society for the Promotion of Science through Grant No. JP19K23424 and from JST FOREST Program (Grant No. JPMJFR222U, Japan).

\bibliography{GZP_references} 

%merlin.mbs apsrev4-1.bst 2010-07-25 4.21a (PWD, AO, DPC) hacked
%Control: key (0)
%Control: author (8) initials jnrlst
%Control: editor formatted (1) identically to author
%Control: production of article title (-1) disabled
%Control: page (0) single
%Control: year (1) truncated
%Control: production of eprint (0) enabled
\begin{thebibliography}{85}%
\makeatletter
\providecommand \@ifxundefined [1]{%
 \@ifx{#1\undefined}
}%
\providecommand \@ifnum [1]{%
 \ifnum #1\expandafter \@firstoftwo
 \else \expandafter \@secondoftwo
 \fi
}%
\providecommand \@ifx [1]{%
 \ifx #1\expandafter \@firstoftwo
 \else \expandafter \@secondoftwo
 \fi
}%
\providecommand \natexlab [1]{#1}%
\providecommand \enquote  [1]{``#1''}%
\providecommand \bibnamefont  [1]{#1}%
\providecommand \bibfnamefont [1]{#1}%
\providecommand \citenamefont [1]{#1}%
\providecommand \href@noop [0]{\@secondoftwo}%
\providecommand \href [0]{\begingroup \@sanitize@url \@href}%
\providecommand \@href[1]{\@@startlink{#1}\@@href}%
\providecommand \@@href[1]{\endgroup#1\@@endlink}%
\providecommand \@sanitize@url [0]{\catcode `\\12\catcode `\$12\catcode
  `\&12\catcode `\#12\catcode `\^12\catcode `\_12\catcode `\%12\relax}%
\providecommand \@@startlink[1]{}%
\providecommand \@@endlink[0]{}%
\providecommand \url  [0]{\begingroup\@sanitize@url \@url }%
\providecommand \@url [1]{\endgroup\@href {#1}{\urlprefix }}%
\providecommand \urlprefix  [0]{URL }%
\providecommand \Eprint [0]{\href }%
\providecommand \doibase [0]{http://dx.doi.org/}%
\providecommand \selectlanguage [0]{\@gobble}%
\providecommand \bibinfo  [0]{\@secondoftwo}%
\providecommand \bibfield  [0]{\@secondoftwo}%
\providecommand \translation [1]{[#1]}%
\providecommand \BibitemOpen [0]{}%
\providecommand \bibitemStop [0]{}%
\providecommand \bibitemNoStop [0]{.\EOS\space}%
\providecommand \EOS [0]{\spacefactor3000\relax}%
\providecommand \BibitemShut  [1]{\csname bibitem#1\endcsname}%
\let\auto@bib@innerbib\@empty
%</preamble>
\bibitem [{\citenamefont {Horodecki}\ \emph {et~al.}(2009)\citenamefont
  {Horodecki}, \citenamefont {Horodecki}, \citenamefont {Horodecki},\ and\
  \citenamefont {Horodecki}}]{RH09}%
  \BibitemOpen
  \bibfield  {author} {\bibinfo {author} {\bibfnamefont {R.}~\bibnamefont
  {Horodecki}}, \bibinfo {author} {\bibfnamefont {P.}~\bibnamefont
  {Horodecki}}, \bibinfo {author} {\bibfnamefont {M.}~\bibnamefont
  {Horodecki}}, \ and\ \bibinfo {author} {\bibfnamefont {K.}~\bibnamefont
  {Horodecki}},\ }\href {\doibase 10.1103/RevModPhys.81.865} {\bibfield
  {journal} {\bibinfo  {journal} {Rev. Mod. Phys.}\ }\textbf {\bibinfo {volume}
  {81}},\ \bibinfo {pages} {865} (\bibinfo {year} {2009})}\BibitemShut
  {NoStop}%
\bibitem [{\citenamefont {Amico}\ \emph {et~al.}(2008)\citenamefont {Amico},
  \citenamefont {Fazio}, \citenamefont {Osterloh},\ and\ \citenamefont
  {Vedral}}]{LA08}%
  \BibitemOpen
  \bibfield  {author} {\bibinfo {author} {\bibfnamefont {L.}~\bibnamefont
  {Amico}}, \bibinfo {author} {\bibfnamefont {R.}~\bibnamefont {Fazio}},
  \bibinfo {author} {\bibfnamefont {A.}~\bibnamefont {Osterloh}}, \ and\
  \bibinfo {author} {\bibfnamefont {V.}~\bibnamefont {Vedral}},\ }\href
  {\doibase 10.1103/RevModPhys.80.517} {\bibfield  {journal} {\bibinfo
  {journal} {Rev. Mod. Phys.}\ }\textbf {\bibinfo {volume} {80}},\ \bibinfo
  {pages} {517} (\bibinfo {year} {2008})}\BibitemShut {NoStop}%
\bibitem [{\citenamefont {Eisert}\ \emph {et~al.}(2010)\citenamefont {Eisert},
  \citenamefont {Cramer},\ and\ \citenamefont {Plenio}}]{JE10}%
  \BibitemOpen
  \bibfield  {author} {\bibinfo {author} {\bibfnamefont {J.}~\bibnamefont
  {Eisert}}, \bibinfo {author} {\bibfnamefont {M.}~\bibnamefont {Cramer}}, \
  and\ \bibinfo {author} {\bibfnamefont {M.~B.}\ \bibnamefont {Plenio}},\
  }\href {\doibase 10.1103/RevModPhys.82.277} {\bibfield  {journal} {\bibinfo
  {journal} {Rev. Mod. Phys.}\ }\textbf {\bibinfo {volume} {82}},\ \bibinfo
  {pages} {277} (\bibinfo {year} {2010})}\BibitemShut {NoStop}%
\bibitem [{\citenamefont {Islam}\ \emph {et~al.}(2015)\citenamefont {Islam},
  \citenamefont {Ma}, \citenamefont {Preiss}, \citenamefont {Tai},
  \citenamefont {Lukin}, \citenamefont {Rispoli},\ and\ \citenamefont
  {Greiner}}]{RI15}%
  \BibitemOpen
  \bibfield  {author} {\bibinfo {author} {\bibfnamefont {R.}~\bibnamefont
  {Islam}}, \bibinfo {author} {\bibfnamefont {R.}~\bibnamefont {Ma}}, \bibinfo
  {author} {\bibfnamefont {P.~M.}\ \bibnamefont {Preiss}}, \bibinfo {author}
  {\bibfnamefont {M.~E.}\ \bibnamefont {Tai}}, \bibinfo {author} {\bibfnamefont
  {A.}~\bibnamefont {Lukin}}, \bibinfo {author} {\bibfnamefont
  {M.}~\bibnamefont {Rispoli}}, \ and\ \bibinfo {author} {\bibfnamefont
  {M.}~\bibnamefont {Greiner}},\ }\href {http://dx.doi.org/10.1038/nature15750}
  {\bibfield  {journal} {\bibinfo  {journal} {Nature}\ }\textbf {\bibinfo
  {volume} {528}},\ \bibinfo {pages} {77} (\bibinfo {year} {2015})}\BibitemShut
  {NoStop}%
\bibitem [{\citenamefont {Laflorencie}(2016)}]{NL16}%
  \BibitemOpen
  \bibfield  {author} {\bibinfo {author} {\bibfnamefont {N.}~\bibnamefont
  {Laflorencie}},\ }\href {\doibase
  https://doi.org/10.1016/j.physrep.2016.06.008} {\bibfield  {journal}
  {\bibinfo  {journal} {Phys. Rep.}\ }\textbf {\bibinfo {volume} {646}},\
  \bibinfo {pages} {1} (\bibinfo {year} {2016})}\BibitemShut {NoStop}%
\bibitem [{\citenamefont {Abanin}\ \emph {et~al.}(2019)\citenamefont {Abanin},
  \citenamefont {Altman}, \citenamefont {Bloch},\ and\ \citenamefont
  {Serbyn}}]{DAA19}%
  \BibitemOpen
  \bibfield  {author} {\bibinfo {author} {\bibfnamefont {D.~A.}\ \bibnamefont
  {Abanin}}, \bibinfo {author} {\bibfnamefont {E.}~\bibnamefont {Altman}},
  \bibinfo {author} {\bibfnamefont {I.}~\bibnamefont {Bloch}}, \ and\ \bibinfo
  {author} {\bibfnamefont {M.}~\bibnamefont {Serbyn}},\ }\href {\doibase
  10.1103/RevModPhys.91.021001} {\bibfield  {journal} {\bibinfo  {journal}
  {Rev. Mod. Phys.}\ }\textbf {\bibinfo {volume} {91}},\ \bibinfo {pages}
  {021001} (\bibinfo {year} {2019})}\BibitemShut {NoStop}%
\bibitem [{\citenamefont {Chiu}\ \emph {et~al.}(2016)\citenamefont {Chiu},
  \citenamefont {Teo}, \citenamefont {Schnyder},\ and\ \citenamefont
  {Ryu}}]{CKC16}%
  \BibitemOpen
  \bibfield  {author} {\bibinfo {author} {\bibfnamefont {C.-K.}\ \bibnamefont
  {Chiu}}, \bibinfo {author} {\bibfnamefont {J.~C.~Y.}\ \bibnamefont {Teo}},
  \bibinfo {author} {\bibfnamefont {A.~P.}\ \bibnamefont {Schnyder}}, \ and\
  \bibinfo {author} {\bibfnamefont {S.}~\bibnamefont {Ryu}},\ }\href {\doibase
  10.1103/RevModPhys.88.035005} {\bibfield  {journal} {\bibinfo  {journal}
  {Rev. Mod. Phys.}\ }\textbf {\bibinfo {volume} {88}},\ \bibinfo {pages}
  {035005} (\bibinfo {year} {2016})}\BibitemShut {NoStop}%
\bibitem [{\citenamefont {Wen}(2017)}]{XGW17}%
  \BibitemOpen
  \bibfield  {author} {\bibinfo {author} {\bibfnamefont {X.-G.}\ \bibnamefont
  {Wen}},\ }\href {\doibase 10.1103/RevModPhys.89.041004} {\bibfield  {journal}
  {\bibinfo  {journal} {Rev. Mod. Phys.}\ }\textbf {\bibinfo {volume} {89}},\
  \bibinfo {pages} {041004} (\bibinfo {year} {2017})}\BibitemShut {NoStop}%
\bibitem [{\citenamefont {Zeng}\ \emph {et~al.}(2019)\citenamefont {Zeng},
  \citenamefont {Chen}, \citenamefont {Zhou},\ and\ \citenamefont
  {Wen}}]{BZ19}%
  \BibitemOpen
  \bibfield  {author} {\bibinfo {author} {\bibfnamefont {B.}~\bibnamefont
  {Zeng}}, \bibinfo {author} {\bibfnamefont {X.}~\bibnamefont {Chen}}, \bibinfo
  {author} {\bibfnamefont {D.-L.}\ \bibnamefont {Zhou}}, \ and\ \bibinfo
  {author} {\bibfnamefont {X.-G.}\ \bibnamefont {Wen}},\ }\href@noop {} {\emph
  {\bibinfo {title} {Quantum information meets quantum matter}}}\ (\bibinfo
  {publisher} {Springer, New York},\ \bibinfo {year} {2019})\BibitemShut
  {NoStop}%
\bibitem [{\citenamefont {Nandkishore}\ and\ \citenamefont
  {Huse}(2015)}]{RN15}%
  \BibitemOpen
  \bibfield  {author} {\bibinfo {author} {\bibfnamefont {R.}~\bibnamefont
  {Nandkishore}}\ and\ \bibinfo {author} {\bibfnamefont {D.~A.}\ \bibnamefont
  {Huse}},\ }\href {https://doi.org/10.1146/annurev-conmatphys-031214-014726}
  {\bibfield  {journal} {\bibinfo  {journal} {Annu. Rev. Cond. Matt. Phys.}\
  }\textbf {\bibinfo {volume} {6}},\ \bibinfo {pages} {201} (\bibinfo {year}
  {2015})}\BibitemShut {NoStop}%
\bibitem [{\citenamefont {Kaufman}\ \emph {et~al.}(2016)\citenamefont
  {Kaufman}, \citenamefont {Tai}, \citenamefont {Lukin}, \citenamefont
  {Rispoli}, \citenamefont {Schittko}, \citenamefont {Preiss},\ and\
  \citenamefont {Greiner}}]{AMK16}%
  \BibitemOpen
  \bibfield  {author} {\bibinfo {author} {\bibfnamefont {A.~M.}\ \bibnamefont
  {Kaufman}}, \bibinfo {author} {\bibfnamefont {M.~E.}\ \bibnamefont {Tai}},
  \bibinfo {author} {\bibfnamefont {A.}~\bibnamefont {Lukin}}, \bibinfo
  {author} {\bibfnamefont {M.}~\bibnamefont {Rispoli}}, \bibinfo {author}
  {\bibfnamefont {R.}~\bibnamefont {Schittko}}, \bibinfo {author}
  {\bibfnamefont {P.~M.}\ \bibnamefont {Preiss}}, \ and\ \bibinfo {author}
  {\bibfnamefont {M.}~\bibnamefont {Greiner}},\ }\href {\doibase
  10.1126/science.aaf6725} {\bibfield  {journal} {\bibinfo  {journal}
  {Science}\ }\textbf {\bibinfo {volume} {353}},\ \bibinfo {pages} {794}
  (\bibinfo {year} {2016})}\BibitemShut {NoStop}%
\bibitem [{\citenamefont {Lukin}\ \emph {et~al.}(2019)\citenamefont {Lukin},
  \citenamefont {Rispoli}, \citenamefont {Schittko}, \citenamefont {Tai},
  \citenamefont {Kaufman}, \citenamefont {Choi}, \citenamefont {Khemani},
  \citenamefont {L{\'e}onard},\ and\ \citenamefont {Greiner}}]{AL19}%
  \BibitemOpen
  \bibfield  {author} {\bibinfo {author} {\bibfnamefont {A.}~\bibnamefont
  {Lukin}}, \bibinfo {author} {\bibfnamefont {M.}~\bibnamefont {Rispoli}},
  \bibinfo {author} {\bibfnamefont {R.}~\bibnamefont {Schittko}}, \bibinfo
  {author} {\bibfnamefont {M.~E.}\ \bibnamefont {Tai}}, \bibinfo {author}
  {\bibfnamefont {A.~M.}\ \bibnamefont {Kaufman}}, \bibinfo {author}
  {\bibfnamefont {S.}~\bibnamefont {Choi}}, \bibinfo {author} {\bibfnamefont
  {V.}~\bibnamefont {Khemani}}, \bibinfo {author} {\bibfnamefont
  {J.}~\bibnamefont {L{\'e}onard}}, \ and\ \bibinfo {author} {\bibfnamefont
  {M.}~\bibnamefont {Greiner}},\ }\href {\doibase 10.1126/science.aau0818}
  {\bibfield  {journal} {\bibinfo  {journal} {Science}\ }\textbf {\bibinfo
  {volume} {364}},\ \bibinfo {pages} {256} (\bibinfo {year}
  {2019})}\BibitemShut {NoStop}%
\bibitem [{\citenamefont {Fisher}\ \emph {et~al.}(2023)\citenamefont {Fisher},
  \citenamefont {Khemani}, \citenamefont {Nahum},\ and\ \citenamefont
  {Vijay}}]{MPAF23}%
  \BibitemOpen
  \bibfield  {author} {\bibinfo {author} {\bibfnamefont {M.~P.}\ \bibnamefont
  {Fisher}}, \bibinfo {author} {\bibfnamefont {V.}~\bibnamefont {Khemani}},
  \bibinfo {author} {\bibfnamefont {A.}~\bibnamefont {Nahum}}, \ and\ \bibinfo
  {author} {\bibfnamefont {S.}~\bibnamefont {Vijay}},\ }\href {\doibase
  https://doi.org/10.1146/annurev-conmatphys-031720-030658} {\bibfield
  {journal} {\bibinfo  {journal} {Annu. Rev. Cond. Matt. Phys.}\ }\textbf
  {\bibinfo {volume} {14}},\ \bibinfo {pages} {335} (\bibinfo {year}
  {2023})}\BibitemShut {NoStop}%
\bibitem [{\citenamefont {Verstraete}\ \emph {et~al.}(2008)\citenamefont
  {Verstraete}, \citenamefont {Murg},\ and\ \citenamefont {Cirac}}]{FV08}%
  \BibitemOpen
  \bibfield  {author} {\bibinfo {author} {\bibfnamefont {F.}~\bibnamefont
  {Verstraete}}, \bibinfo {author} {\bibfnamefont {V.}~\bibnamefont {Murg}}, \
  and\ \bibinfo {author} {\bibfnamefont {J.~I.}\ \bibnamefont {Cirac}},\ }\href
  {\doibase 10.1080/14789940801912366} {\bibfield  {journal} {\bibinfo
  {journal} {Adv. Phys.}\ }\textbf {\bibinfo {volume} {57}},\ \bibinfo {pages}
  {143} (\bibinfo {year} {2008})}\BibitemShut {NoStop}%
\bibitem [{\citenamefont {Schollw\"ock}(2011)}]{US11}%
  \BibitemOpen
  \bibfield  {author} {\bibinfo {author} {\bibfnamefont {U.}~\bibnamefont
  {Schollw\"ock}},\ }\href {\doibase https://doi.org/10.1016/j.aop.2010.09.012}
  {\bibfield  {journal} {\bibinfo  {journal} {Ann. Phys.}\ }\textbf {\bibinfo
  {volume} {326}},\ \bibinfo {pages} {96} (\bibinfo {year} {2011})}\BibitemShut
  {NoStop}%
\bibitem [{\citenamefont {Cirac}\ \emph {et~al.}(2021)\citenamefont {Cirac},
  \citenamefont {P\'erez-Garc\'{\i}a}, \citenamefont {Schuch},\ and\
  \citenamefont {Verstraete}}]{JIC21}%
  \BibitemOpen
  \bibfield  {author} {\bibinfo {author} {\bibfnamefont {J.~I.}\ \bibnamefont
  {Cirac}}, \bibinfo {author} {\bibfnamefont {D.}~\bibnamefont
  {P\'erez-Garc\'{\i}a}}, \bibinfo {author} {\bibfnamefont {N.}~\bibnamefont
  {Schuch}}, \ and\ \bibinfo {author} {\bibfnamefont {F.}~\bibnamefont
  {Verstraete}},\ }\href {\doibase 10.1103/RevModPhys.93.045003} {\bibfield
  {journal} {\bibinfo  {journal} {Rev. Mod. Phys.}\ }\textbf {\bibinfo {volume}
  {93}},\ \bibinfo {pages} {045003} (\bibinfo {year} {2021})}\BibitemShut
  {NoStop}%
\bibitem [{\citenamefont {Aolita}\ \emph {et~al.}(2015)\citenamefont {Aolita},
  \citenamefont {de~Melo},\ and\ \citenamefont {Davidovich}}]{LA15}%
  \BibitemOpen
  \bibfield  {author} {\bibinfo {author} {\bibfnamefont {L.}~\bibnamefont
  {Aolita}}, \bibinfo {author} {\bibfnamefont {F.}~\bibnamefont {de~Melo}}, \
  and\ \bibinfo {author} {\bibfnamefont {L.}~\bibnamefont {Davidovich}},\
  }\href {\doibase 10.1088/0034-4885/78/4/042001} {\bibfield  {journal}
  {\bibinfo  {journal} {Rep. Prog. Phys.}\ }\textbf {\bibinfo {volume} {78}},\
  \bibinfo {pages} {042001} (\bibinfo {year} {2015})}\BibitemShut {NoStop}%
\bibitem [{\citenamefont {Yu}\ and\ \citenamefont {Eberly}(2004)}]{TY04}%
  \BibitemOpen
  \bibfield  {author} {\bibinfo {author} {\bibfnamefont {T.}~\bibnamefont
  {Yu}}\ and\ \bibinfo {author} {\bibfnamefont {J.~H.}\ \bibnamefont
  {Eberly}},\ }\href {\doibase 10.1103/PhysRevLett.93.140404} {\bibfield
  {journal} {\bibinfo  {journal} {Phys. Rev. Lett.}\ }\textbf {\bibinfo
  {volume} {93}},\ \bibinfo {pages} {140404} (\bibinfo {year}
  {2004})}\BibitemShut {NoStop}%
\bibitem [{\citenamefont {Yu}\ and\ \citenamefont {Eberly}(2006)}]{TY06}%
  \BibitemOpen
  \bibfield  {author} {\bibinfo {author} {\bibfnamefont {T.}~\bibnamefont
  {Yu}}\ and\ \bibinfo {author} {\bibfnamefont {J.~H.}\ \bibnamefont
  {Eberly}},\ }\href {\doibase 10.1103/PhysRevLett.97.140403} {\bibfield
  {journal} {\bibinfo  {journal} {Phys. Rev. Lett.}\ }\textbf {\bibinfo
  {volume} {97}},\ \bibinfo {pages} {140403} (\bibinfo {year}
  {2006})}\BibitemShut {NoStop}%
\bibitem [{\citenamefont {Bellomo}\ \emph {et~al.}(2007)\citenamefont
  {Bellomo}, \citenamefont {Lo~Franco},\ and\ \citenamefont {Compagno}}]{BB07}%
  \BibitemOpen
  \bibfield  {author} {\bibinfo {author} {\bibfnamefont {B.}~\bibnamefont
  {Bellomo}}, \bibinfo {author} {\bibfnamefont {R.}~\bibnamefont {Lo~Franco}},
  \ and\ \bibinfo {author} {\bibfnamefont {G.}~\bibnamefont {Compagno}},\
  }\href {\doibase 10.1103/PhysRevLett.99.160502} {\bibfield  {journal}
  {\bibinfo  {journal} {Phys. Rev. Lett.}\ }\textbf {\bibinfo {volume} {99}},\
  \bibinfo {pages} {160502} (\bibinfo {year} {2007})}\BibitemShut {NoStop}%
\bibitem [{\citenamefont {Mazzola}\ \emph {et~al.}(2010)\citenamefont
  {Mazzola}, \citenamefont {Piilo},\ and\ \citenamefont {Maniscalco}}]{LM10}%
  \BibitemOpen
  \bibfield  {author} {\bibinfo {author} {\bibfnamefont {L.}~\bibnamefont
  {Mazzola}}, \bibinfo {author} {\bibfnamefont {J.}~\bibnamefont {Piilo}}, \
  and\ \bibinfo {author} {\bibfnamefont {S.}~\bibnamefont {Maniscalco}},\
  }\href {\doibase 10.1103/PhysRevLett.104.200401} {\bibfield  {journal}
  {\bibinfo  {journal} {Phys. Rev. Lett.}\ }\textbf {\bibinfo {volume} {104}},\
  \bibinfo {pages} {200401} (\bibinfo {year} {2010})}\BibitemShut {NoStop}%
\bibitem [{\citenamefont {Yu}\ and\ \citenamefont {Eberly}(2009)}]{TY09}%
  \BibitemOpen
  \bibfield  {author} {\bibinfo {author} {\bibfnamefont {T.}~\bibnamefont
  {Yu}}\ and\ \bibinfo {author} {\bibfnamefont {J.~H.}\ \bibnamefont
  {Eberly}},\ }\href {\doibase 10.1126/science.1167343} {\bibfield  {journal}
  {\bibinfo  {journal} {Science}\ }\textbf {\bibinfo {volume} {323}},\ \bibinfo
  {pages} {598} (\bibinfo {year} {2009})}\BibitemShut {NoStop}%
\bibitem [{\citenamefont {Simon}\ and\ \citenamefont {Kempe}(2002)}]{CS02}%
  \BibitemOpen
  \bibfield  {author} {\bibinfo {author} {\bibfnamefont {C.}~\bibnamefont
  {Simon}}\ and\ \bibinfo {author} {\bibfnamefont {J.}~\bibnamefont {Kempe}},\
  }\href {\doibase 10.1103/PhysRevA.65.052327} {\bibfield  {journal} {\bibinfo
  {journal} {Phys. Rev. A}\ }\textbf {\bibinfo {volume} {65}},\ \bibinfo
  {pages} {052327} (\bibinfo {year} {2002})}\BibitemShut {NoStop}%
\bibitem [{\citenamefont {D\"ur}\ and\ \citenamefont {Briegel}(2004)}]{WD04}%
  \BibitemOpen
  \bibfield  {author} {\bibinfo {author} {\bibfnamefont {W.}~\bibnamefont
  {D\"ur}}\ and\ \bibinfo {author} {\bibfnamefont {H.-J.}\ \bibnamefont
  {Briegel}},\ }\href {\doibase 10.1103/PhysRevLett.92.180403} {\bibfield
  {journal} {\bibinfo  {journal} {Phys. Rev. Lett.}\ }\textbf {\bibinfo
  {volume} {92}},\ \bibinfo {pages} {180403} (\bibinfo {year}
  {2004})}\BibitemShut {NoStop}%
\bibitem [{\citenamefont {Hein}\ \emph {et~al.}(2005)\citenamefont {Hein},
  \citenamefont {D\"ur},\ and\ \citenamefont {Briegel}}]{MH05}%
  \BibitemOpen
  \bibfield  {author} {\bibinfo {author} {\bibfnamefont {M.}~\bibnamefont
  {Hein}}, \bibinfo {author} {\bibfnamefont {W.}~\bibnamefont {D\"ur}}, \ and\
  \bibinfo {author} {\bibfnamefont {H.-J.}\ \bibnamefont {Briegel}},\ }\href
  {\doibase 10.1103/PhysRevA.71.032350} {\bibfield  {journal} {\bibinfo
  {journal} {Phys. Rev. A}\ }\textbf {\bibinfo {volume} {71}},\ \bibinfo
  {pages} {032350} (\bibinfo {year} {2005})}\BibitemShut {NoStop}%
\bibitem [{\citenamefont {Aolita}\ \emph {et~al.}(2008)\citenamefont {Aolita},
  \citenamefont {Chaves}, \citenamefont {Cavalcanti}, \citenamefont
  {Ac\'{\i}n},\ and\ \citenamefont {Davidovich}}]{LD08}%
  \BibitemOpen
  \bibfield  {author} {\bibinfo {author} {\bibfnamefont {L.}~\bibnamefont
  {Aolita}}, \bibinfo {author} {\bibfnamefont {R.}~\bibnamefont {Chaves}},
  \bibinfo {author} {\bibfnamefont {D.}~\bibnamefont {Cavalcanti}}, \bibinfo
  {author} {\bibfnamefont {A.}~\bibnamefont {Ac\'{\i}n}}, \ and\ \bibinfo
  {author} {\bibfnamefont {L.}~\bibnamefont {Davidovich}},\ }\href {\doibase
  10.1103/PhysRevLett.100.080501} {\bibfield  {journal} {\bibinfo  {journal}
  {Phys. Rev. Lett.}\ }\textbf {\bibinfo {volume} {100}},\ \bibinfo {pages}
  {080501} (\bibinfo {year} {2008})}\BibitemShut {NoStop}%
\bibitem [{\citenamefont {Wang}\ \emph {et~al.}(2010)\citenamefont {Wang},
  \citenamefont {Miranowicz}, \citenamefont {Liu}, \citenamefont {Sun},\ and\
  \citenamefont {Nori}}]{XW10}%
  \BibitemOpen
  \bibfield  {author} {\bibinfo {author} {\bibfnamefont {X.}~\bibnamefont
  {Wang}}, \bibinfo {author} {\bibfnamefont {A.}~\bibnamefont {Miranowicz}},
  \bibinfo {author} {\bibfnamefont {Y.-x.}\ \bibnamefont {Liu}}, \bibinfo
  {author} {\bibfnamefont {C.~P.}\ \bibnamefont {Sun}}, \ and\ \bibinfo
  {author} {\bibfnamefont {F.}~\bibnamefont {Nori}},\ }\href {\doibase
  10.1103/PhysRevA.81.022106} {\bibfield  {journal} {\bibinfo  {journal} {Phys.
  Rev. A}\ }\textbf {\bibinfo {volume} {81}},\ \bibinfo {pages} {022106}
  (\bibinfo {year} {2010})}\BibitemShut {NoStop}%
\bibitem [{\citenamefont {Vedral}\ \emph {et~al.}(1997)\citenamefont {Vedral},
  \citenamefont {Plenio}, \citenamefont {Rippin},\ and\ \citenamefont
  {Knight}}]{VV97}%
  \BibitemOpen
  \bibfield  {author} {\bibinfo {author} {\bibfnamefont {V.}~\bibnamefont
  {Vedral}}, \bibinfo {author} {\bibfnamefont {M.~B.}\ \bibnamefont {Plenio}},
  \bibinfo {author} {\bibfnamefont {M.~A.}\ \bibnamefont {Rippin}}, \ and\
  \bibinfo {author} {\bibfnamefont {P.~L.}\ \bibnamefont {Knight}},\ }\href
  {\doibase 10.1103/PhysRevLett.78.2275} {\bibfield  {journal} {\bibinfo
  {journal} {Phys. Rev. Lett.}\ }\textbf {\bibinfo {volume} {78}},\ \bibinfo
  {pages} {2275} (\bibinfo {year} {1997})}\BibitemShut {NoStop}%
\bibitem [{\citenamefont {S\'a}\ \emph {et~al.}(2020)\citenamefont {S\'a},
  \citenamefont {Ribeiro}, \citenamefont {Can},\ and\ \citenamefont
  {Prosen}}]{LS20}%
  \BibitemOpen
  \bibfield  {author} {\bibinfo {author} {\bibfnamefont {L.}~\bibnamefont
  {S\'a}}, \bibinfo {author} {\bibfnamefont {P.}~\bibnamefont {Ribeiro}},
  \bibinfo {author} {\bibfnamefont {T.}~\bibnamefont {Can}}, \ and\ \bibinfo
  {author} {\bibfnamefont {T.}~\bibnamefont {Prosen}},\ }\href {\doibase
  10.1103/PhysRevB.102.134310} {\bibfield  {journal} {\bibinfo  {journal}
  {Phys. Rev. B}\ }\textbf {\bibinfo {volume} {102}},\ \bibinfo {pages}
  {134310} (\bibinfo {year} {2020})}\BibitemShut {NoStop}%
\bibitem [{\citenamefont {Wolf}\ and\ \citenamefont {Cirac}(2008)}]{JIC08}%
  \BibitemOpen
  \bibfield  {author} {\bibinfo {author} {\bibfnamefont {M.~M.}\ \bibnamefont
  {Wolf}}\ and\ \bibinfo {author} {\bibfnamefont {J.~I.}\ \bibnamefont
  {Cirac}},\ }\href {https://doi.org/10.1007/s00220-008-0411-y} {\bibfield
  {journal} {\bibinfo  {journal} {Commun. Math. Phys.}\ }\textbf {\bibinfo
  {volume} {279}},\ \bibinfo {pages} {147} (\bibinfo {year}
  {2008})}\BibitemShut {NoStop}%
\bibitem [{\citenamefont {Plenio}\ and\ \citenamefont {Virmani}(2007)}]{MBP07}%
  \BibitemOpen
  \bibfield  {author} {\bibinfo {author} {\bibfnamefont {M.~B.}\ \bibnamefont
  {Plenio}}\ and\ \bibinfo {author} {\bibfnamefont {S.}~\bibnamefont
  {Virmani}},\ }\href {\doibase 10.26421/QIC7.1-2-1} {\bibfield  {journal}
  {\bibinfo  {journal} {Quantum Info. Comput.}\ }\textbf {\bibinfo {volume}
  {7}},\ \bibinfo {pages} {1} (\bibinfo {year} {2007})}\BibitemShut {NoStop}%
\bibitem [{\citenamefont {Walter}\ \emph {et~al.}(2016)\citenamefont {Walter},
  \citenamefont {Gross},\ and\ \citenamefont {Eisert}}]{MW16}%
  \BibitemOpen
  \bibfield  {author} {\bibinfo {author} {\bibfnamefont {M.}~\bibnamefont
  {Walter}}, \bibinfo {author} {\bibfnamefont {D.}~\bibnamefont {Gross}}, \
  and\ \bibinfo {author} {\bibfnamefont {J.}~\bibnamefont {Eisert}},\ }\enquote
  {\bibinfo {title} {Multi-partite entanglement},}\ \ (\bibinfo  {publisher}
  {Wiley-VCH},\ \bibinfo {address} {Hoboken, New Jersey},\ \bibinfo {year}
  {2016})\ pp.\ \bibinfo {pages} {293--330}\BibitemShut {NoStop}%
\bibitem [{\citenamefont {Horodecki}\ \emph {et~al.}(2024)\citenamefont
  {Horodecki}, \citenamefont {Rudnicki},\ and\ \citenamefont
  {\.Zyczkowski}}]{PH24}%
  \BibitemOpen
  \bibfield  {author} {\bibinfo {author} {\bibfnamefont {P.}~\bibnamefont
  {Horodecki}}, \bibinfo {author} {\bibfnamefont {L.}~\bibnamefont {Rudnicki}},
  \ and\ \bibinfo {author} {\bibfnamefont {K.}~\bibnamefont {\.Zyczkowski}},\
  }\href {https://arxiv.org/abs/2409.04566} {\enquote {\bibinfo {title}
  {Multipartite entanglement},}\ } (\bibinfo {year} {2024}),\ \bibinfo {note}
  {arXiv:2409.04566}\BibitemShut {NoStop}%
\bibitem [{Note1()}]{Note1}%
  \BibitemOpen
  \bibinfo {note} {One can adopt other norms, such as the trace norm more
  commonly used in the literature \cite {MJK12,OS15}. However, since we focus
  on a \protect \emph {local} Hilbert space of finite dimension, different
  norms only differ by a constant factor independent of $N$ \cite {RAH12}, thus
  the convergence bound (\ref {cb}) will have no formal
  difference.}\BibitemShut {Stop}%
\bibitem [{\citenamefont {Kitaev}\ and\ \citenamefont {Preskill}(2006)}]{AK06}%
  \BibitemOpen
  \bibfield  {author} {\bibinfo {author} {\bibfnamefont {A.}~\bibnamefont
  {Kitaev}}\ and\ \bibinfo {author} {\bibfnamefont {J.}~\bibnamefont
  {Preskill}},\ }\href {\doibase 10.1103/PhysRevLett.96.110404} {\bibfield
  {journal} {\bibinfo  {journal} {Phys. Rev. Lett.}\ }\textbf {\bibinfo
  {volume} {96}},\ \bibinfo {pages} {110404} (\bibinfo {year}
  {2006})}\BibitemShut {NoStop}%
\bibitem [{\citenamefont {Levin}\ and\ \citenamefont {Wen}(2006)}]{ML06}%
  \BibitemOpen
  \bibfield  {author} {\bibinfo {author} {\bibfnamefont {M.}~\bibnamefont
  {Levin}}\ and\ \bibinfo {author} {\bibfnamefont {X.-G.}\ \bibnamefont
  {Wen}},\ }\href {\doibase 10.1103/PhysRevLett.96.110405} {\bibfield
  {journal} {\bibinfo  {journal} {Phys. Rev. Lett.}\ }\textbf {\bibinfo
  {volume} {96}},\ \bibinfo {pages} {110405} (\bibinfo {year}
  {2006})}\BibitemShut {NoStop}%
\bibitem [{\citenamefont {Jiang}\ \emph {et~al.}(2012)\citenamefont {Jiang},
  \citenamefont {Wang},\ and\ \citenamefont {Balents}}]{HCJ12}%
  \BibitemOpen
  \bibfield  {author} {\bibinfo {author} {\bibfnamefont {H.-C.}\ \bibnamefont
  {Jiang}}, \bibinfo {author} {\bibfnamefont {Z.}~\bibnamefont {Wang}}, \ and\
  \bibinfo {author} {\bibfnamefont {L.}~\bibnamefont {Balents}},\ }\href
  {\doibase 10.1038/nphys2465} {\bibfield  {journal} {\bibinfo  {journal} {Nat.
  Phys.}\ }\textbf {\bibinfo {volume} {8}},\ \bibinfo {pages} {902} (\bibinfo
  {year} {2012})}\BibitemShut {NoStop}%
\bibitem [{\citenamefont {Page}(1993)}]{DNP93}%
  \BibitemOpen
  \bibfield  {author} {\bibinfo {author} {\bibfnamefont {D.~N.}\ \bibnamefont
  {Page}},\ }\href {\doibase 10.1103/PhysRevLett.71.1291} {\bibfield  {journal}
  {\bibinfo  {journal} {Phys. Rev. Lett.}\ }\textbf {\bibinfo {volume} {71}},\
  \bibinfo {pages} {1291} (\bibinfo {year} {1993})}\BibitemShut {NoStop}%
\bibitem [{\citenamefont {Popescu}\ \emph {et~al.}(2006)\citenamefont
  {Popescu}, \citenamefont {Short},\ and\ \citenamefont {Winter}}]{SP06}%
  \BibitemOpen
  \bibfield  {author} {\bibinfo {author} {\bibfnamefont {S.}~\bibnamefont
  {Popescu}}, \bibinfo {author} {\bibfnamefont {A.~J.}\ \bibnamefont {Short}},
  \ and\ \bibinfo {author} {\bibfnamefont {A.}~\bibnamefont {Winter}},\ }\href
  {\doibase 10.1038/nphys444} {\bibfield  {journal} {\bibinfo  {journal} {Nat.
  Phys.}\ }\textbf {\bibinfo {volume} {2}},\ \bibinfo {pages}
  {754^^e2^^80^^93758} (\bibinfo {year} {2006})}\BibitemShut {NoStop}%
\bibitem [{\citenamefont {Bianchi}\ \emph {et~al.}(2022)\citenamefont
  {Bianchi}, \citenamefont {Hackl}, \citenamefont {Kieburg}, \citenamefont
  {Rigol},\ and\ \citenamefont {Vidmar}}]{EB22}%
  \BibitemOpen
  \bibfield  {author} {\bibinfo {author} {\bibfnamefont {E.}~\bibnamefont
  {Bianchi}}, \bibinfo {author} {\bibfnamefont {L.}~\bibnamefont {Hackl}},
  \bibinfo {author} {\bibfnamefont {M.}~\bibnamefont {Kieburg}}, \bibinfo
  {author} {\bibfnamefont {M.}~\bibnamefont {Rigol}}, \ and\ \bibinfo {author}
  {\bibfnamefont {L.}~\bibnamefont {Vidmar}},\ }\href {\doibase
  10.1103/PRXQuantum.3.030201} {\bibfield  {journal} {\bibinfo  {journal} {PRX
  Quantum}\ }\textbf {\bibinfo {volume} {3}},\ \bibinfo {pages} {030201}
  (\bibinfo {year} {2022})}\BibitemShut {NoStop}%
\bibitem [{\citenamefont {Kastoryano}\ \emph {et~al.}(2012)\citenamefont
  {Kastoryano}, \citenamefont {Reeb},\ and\ \citenamefont {Wolf}}]{MJK12}%
  \BibitemOpen
  \bibfield  {author} {\bibinfo {author} {\bibfnamefont {M.~J.}\ \bibnamefont
  {Kastoryano}}, \bibinfo {author} {\bibfnamefont {D.}~\bibnamefont {Reeb}}, \
  and\ \bibinfo {author} {\bibfnamefont {M.~M.}\ \bibnamefont {Wolf}},\ }\href
  {\doibase 10.1088/1751-8113/45/7/075307} {\bibfield  {journal} {\bibinfo
  {journal} {J. Phys. A: Math. Theor.}\ }\textbf {\bibinfo {volume} {45}},\
  \bibinfo {pages} {075307} (\bibinfo {year} {2012})}\BibitemShut {NoStop}%
\bibitem [{\citenamefont {Szehr}\ \emph {et~al.}(2015)\citenamefont {Szehr},
  \citenamefont {Reeb},\ and\ \citenamefont {Wolf}}]{OS15}%
  \BibitemOpen
  \bibfield  {author} {\bibinfo {author} {\bibfnamefont {O.}~\bibnamefont
  {Szehr}}, \bibinfo {author} {\bibfnamefont {D.}~\bibnamefont {Reeb}}, \ and\
  \bibinfo {author} {\bibfnamefont {M.~M.}\ \bibnamefont {Wolf}},\ }\href
  {https://doi.org/10.1007/s00220-014-2188-5} {\bibfield  {journal} {\bibinfo
  {journal} {Commun. Math. Phys.}\ }\textbf {\bibinfo {volume} {333}},\
  \bibinfo {pages} {565} (\bibinfo {year} {2015})}\BibitemShut {NoStop}%
\bibitem [{\citenamefont {Fannes}\ \emph {et~al.}(1992)\citenamefont {Fannes},
  \citenamefont {Nachtergaele},\ and\ \citenamefont {Werner}}]{MF92}%
  \BibitemOpen
  \bibfield  {author} {\bibinfo {author} {\bibfnamefont {M.}~\bibnamefont
  {Fannes}}, \bibinfo {author} {\bibfnamefont {B.}~\bibnamefont
  {Nachtergaele}}, \ and\ \bibinfo {author} {\bibfnamefont {R.~F.}\
  \bibnamefont {Werner}},\ }\href {\doibase 10.1007/BF02099178} {\bibfield
  {journal} {\bibinfo  {journal} {Commun. Math. Phys.}\ }\textbf {\bibinfo
  {volume} {144}},\ \bibinfo {pages} {443} (\bibinfo {year}
  {1992})}\BibitemShut {NoStop}%
\bibitem [{\citenamefont {Weingarten}(1978)}]{DW78}%
  \BibitemOpen
  \bibfield  {author} {\bibinfo {author} {\bibfnamefont {D.}~\bibnamefont
  {Weingarten}},\ }\href {\doibase 10.1063/1.523807} {\bibfield  {journal}
  {\bibinfo  {journal} {J. Math. Phys.}\ }\textbf {\bibinfo {volume} {19}},\
  \bibinfo {pages} {999} (\bibinfo {year} {1978})}\BibitemShut {NoStop}%
\bibitem [{\citenamefont {Brouwer}\ and\ \citenamefont
  {Beenakker}(1996)}]{PWB96}%
  \BibitemOpen
  \bibfield  {author} {\bibinfo {author} {\bibfnamefont {P.~W.}\ \bibnamefont
  {Brouwer}}\ and\ \bibinfo {author} {\bibfnamefont {C.~W.~J.}\ \bibnamefont
  {Beenakker}},\ }\href {\doibase 10.1063/1.531667} {\bibfield  {journal}
  {\bibinfo  {journal} {J. Math. Phys.}\ }\textbf {\bibinfo {volume} {37}},\
  \bibinfo {pages} {4904} (\bibinfo {year} {1996})}\BibitemShut {NoStop}%
\bibitem [{\citenamefont {Collins}\ and\ \citenamefont
  {\'{S}niady}(2006)}]{BC06}%
  \BibitemOpen
  \bibfield  {author} {\bibinfo {author} {\bibfnamefont {B.}~\bibnamefont
  {Collins}}\ and\ \bibinfo {author} {\bibfnamefont {P.}~\bibnamefont
  {\'{S}niady}},\ }\href {\doibase 10.1007/s00220-006-1554-3} {\bibfield
  {journal} {\bibinfo  {journal} {Commun. Math. Phys.}\ }\textbf {\bibinfo
  {volume} {264}},\ \bibinfo {pages} {773} (\bibinfo {year}
  {2006})}\BibitemShut {NoStop}%
\bibitem [{\citenamefont {Horodecki}\ and\ \citenamefont
  {Horodecki}(1999)}]{MH99}%
  \BibitemOpen
  \bibfield  {author} {\bibinfo {author} {\bibfnamefont {M.}~\bibnamefont
  {Horodecki}}\ and\ \bibinfo {author} {\bibfnamefont {P.}~\bibnamefont
  {Horodecki}},\ }\href {\doibase 10.1103/PhysRevA.59.4206} {\bibfield
  {journal} {\bibinfo  {journal} {Phys. Rev. A}\ }\textbf {\bibinfo {volume}
  {59}},\ \bibinfo {pages} {4206} (\bibinfo {year} {1999})}\BibitemShut
  {NoStop}%
\bibitem [{Note2()}]{Note2}%
  \BibitemOpen
  \bibinfo {note} {For the qudit depolarization channel, we have $\|\protect
  \mathcal {E}^t(\protect \hat \rho )-\protect \hat {\protect \mathbb {1}}/d\|=
  p^t\|\protect \hat \rho -\protect \hat {\protect \mathbb {1}}/d\|\le
  (d-1)p^t/d$, so we can take $C=(d-1)/d$ and $\kappa =-\ln p$. Combining with
  $\lambda =1/d$, we obtain $t_{\protect \rm c}=-\ln [(d-1)(d^2+1)]/\ln
  p$.}\BibitemShut {Stop}%
\bibitem [{\citenamefont {Sagawa}\ and\ \citenamefont {Ueda}(2008)}]{TS08}%
  \BibitemOpen
  \bibfield  {author} {\bibinfo {author} {\bibfnamefont {T.}~\bibnamefont
  {Sagawa}}\ and\ \bibinfo {author} {\bibfnamefont {M.}~\bibnamefont {Ueda}},\
  }\href {\doibase 10.1103/PhysRevLett.100.080403} {\bibfield  {journal}
  {\bibinfo  {journal} {Phys. Rev. Lett.}\ }\textbf {\bibinfo {volume} {100}},\
  \bibinfo {pages} {080403} (\bibinfo {year} {2008})}\BibitemShut {NoStop}%
\bibitem [{\citenamefont {Nielsen}\ and\ \citenamefont {Chuang}(2010)}]{MAN10}%
  \BibitemOpen
  \bibfield  {author} {\bibinfo {author} {\bibfnamefont {M.~A.}\ \bibnamefont
  {Nielsen}}\ and\ \bibinfo {author} {\bibfnamefont {I.~L.}\ \bibnamefont
  {Chuang}},\ }\href@noop {} {\emph {\bibinfo {title} {Quantum Computation and
  Information}}}\ (\bibinfo  {publisher} {Cambridge University Press,
  Cambridge},\ \bibinfo {year} {2010})\BibitemShut {NoStop}%
\bibitem [{\citenamefont {Macieszczak}\ \emph {et~al.}(2016)\citenamefont
  {Macieszczak}, \citenamefont {Gu\c{t}\u{a}}, \citenamefont {Lesanovsky},\
  and\ \citenamefont {Garrahan}}]{KM16}%
  \BibitemOpen
  \bibfield  {author} {\bibinfo {author} {\bibfnamefont {K.}~\bibnamefont
  {Macieszczak}}, \bibinfo {author} {\bibfnamefont {M.}~\bibnamefont
  {Gu\c{t}\u{a}}}, \bibinfo {author} {\bibfnamefont {I.}~\bibnamefont
  {Lesanovsky}}, \ and\ \bibinfo {author} {\bibfnamefont {J.~P.}\ \bibnamefont
  {Garrahan}},\ }\href {\doibase 10.1103/PhysRevLett.116.240404} {\bibfield
  {journal} {\bibinfo  {journal} {Phys. Rev. Lett.}\ }\textbf {\bibinfo
  {volume} {116}},\ \bibinfo {pages} {240404} (\bibinfo {year}
  {2016})}\BibitemShut {NoStop}%
\bibitem [{\citenamefont {Heiss}(2012)}]{WDH12}%
  \BibitemOpen
  \bibfield  {author} {\bibinfo {author} {\bibfnamefont {W.~D.}\ \bibnamefont
  {Heiss}},\ }\href {\doibase 10.1088/1751-8113/45/44/444016} {\bibfield
  {journal} {\bibinfo  {journal} {J. Phys. A: Math. Theor.}\ }\textbf {\bibinfo
  {volume} {45}},\ \bibinfo {pages} {444016} (\bibinfo {year}
  {2012})}\BibitemShut {NoStop}%
\bibitem [{\citenamefont {Ashida}\ \emph {et~al.}(2021)\citenamefont {Ashida},
  \citenamefont {Gong},\ and\ \citenamefont {Ueda}}]{YA21}%
  \BibitemOpen
  \bibfield  {author} {\bibinfo {author} {\bibfnamefont {Y.}~\bibnamefont
  {Ashida}}, \bibinfo {author} {\bibfnamefont {Z.}~\bibnamefont {Gong}}, \ and\
  \bibinfo {author} {\bibfnamefont {M.}~\bibnamefont {Ueda}},\ }\href {\doibase
  10.1080/00018732.2021.1876991} {\bibfield  {journal} {\bibinfo  {journal}
  {Adv. Phys.}\ }\textbf {\bibinfo {volume} {69}},\ \bibinfo {pages} {249}
  (\bibinfo {year} {2021})}\BibitemShut {NoStop}%
\bibitem [{\citenamefont {Gross}\ \emph {et~al.}(2007)\citenamefont {Gross},
  \citenamefont {Audenaert},\ and\ \citenamefont {Eisert}}]{DG07}%
  \BibitemOpen
  \bibfield  {author} {\bibinfo {author} {\bibfnamefont {D.}~\bibnamefont
  {Gross}}, \bibinfo {author} {\bibfnamefont {K.}~\bibnamefont {Audenaert}}, \
  and\ \bibinfo {author} {\bibfnamefont {J.}~\bibnamefont {Eisert}},\ }\href
  {\doibase 10.1063/1.2716992} {\bibfield  {journal} {\bibinfo  {journal} {J.
  Math. Phys.}\ }\textbf {\bibinfo {volume} {48}},\ \bibinfo {pages} {052104}
  (\bibinfo {year} {2007})}\BibitemShut {NoStop}%
\bibitem [{\citenamefont {Dankert}\ \emph {et~al.}(2009)\citenamefont
  {Dankert}, \citenamefont {Cleve}, \citenamefont {Emerson},\ and\
  \citenamefont {Livine}}]{CD09}%
  \BibitemOpen
  \bibfield  {author} {\bibinfo {author} {\bibfnamefont {C.}~\bibnamefont
  {Dankert}}, \bibinfo {author} {\bibfnamefont {R.}~\bibnamefont {Cleve}},
  \bibinfo {author} {\bibfnamefont {J.}~\bibnamefont {Emerson}}, \ and\
  \bibinfo {author} {\bibfnamefont {E.}~\bibnamefont {Livine}},\ }\href
  {\doibase 10.1103/PhysRevA.80.012304} {\bibfield  {journal} {\bibinfo
  {journal} {Phys. Rev. A}\ }\textbf {\bibinfo {volume} {80}},\ \bibinfo
  {pages} {012304} (\bibinfo {year} {2009})}\BibitemShut {NoStop}%
\bibitem [{\citenamefont {Renes}\ \emph {et~al.}(2004)\citenamefont {Renes},
  \citenamefont {Blume-Kohout}, \citenamefont {Scott},\ and\ \citenamefont
  {Caves}}]{JMR04}%
  \BibitemOpen
  \bibfield  {author} {\bibinfo {author} {\bibfnamefont {J.~M.}\ \bibnamefont
  {Renes}}, \bibinfo {author} {\bibfnamefont {R.}~\bibnamefont {Blume-Kohout}},
  \bibinfo {author} {\bibfnamefont {A.~J.}\ \bibnamefont {Scott}}, \ and\
  \bibinfo {author} {\bibfnamefont {C.~M.}\ \bibnamefont {Caves}},\ }\href
  {\doibase 10.1063/1.1737053} {\bibfield  {journal} {\bibinfo  {journal} {J.
  Math. Phys.}\ }\textbf {\bibinfo {volume} {45}},\ \bibinfo {pages} {2171}
  (\bibinfo {year} {2004})}\BibitemShut {NoStop}%
\bibitem [{\citenamefont {Huang}\ \emph {et~al.}(2020)\citenamefont {Huang},
  \citenamefont {Kueng},\ and\ \citenamefont {Preskill}}]{HY20}%
  \BibitemOpen
  \bibfield  {author} {\bibinfo {author} {\bibfnamefont {H.-Y.}\ \bibnamefont
  {Huang}}, \bibinfo {author} {\bibfnamefont {R.}~\bibnamefont {Kueng}}, \ and\
  \bibinfo {author} {\bibfnamefont {J.}~\bibnamefont {Preskill}},\ }\href
  {\doibase 10.1038/s41567-020-0932-7} {\bibfield  {journal} {\bibinfo
  {journal} {Nat. Phys.}\ }\textbf {\bibinfo {volume} {16}},\ \bibinfo {pages}
  {1050} (\bibinfo {year} {2020})}\BibitemShut {NoStop}%
\bibitem [{\citenamefont {Chen}\ \emph {et~al.}(2021)\citenamefont {Chen},
  \citenamefont {Yu}, \citenamefont {Zeng},\ and\ \citenamefont
  {Flammia}}]{SC21}%
  \BibitemOpen
  \bibfield  {author} {\bibinfo {author} {\bibfnamefont {S.}~\bibnamefont
  {Chen}}, \bibinfo {author} {\bibfnamefont {W.}~\bibnamefont {Yu}}, \bibinfo
  {author} {\bibfnamefont {P.}~\bibnamefont {Zeng}}, \ and\ \bibinfo {author}
  {\bibfnamefont {S.~T.}\ \bibnamefont {Flammia}},\ }\href {\doibase
  10.1103/PRXQuantum.2.030348} {\bibfield  {journal} {\bibinfo  {journal} {PRX
  Quantum}\ }\textbf {\bibinfo {volume} {2}},\ \bibinfo {pages} {030348}
  (\bibinfo {year} {2021})}\BibitemShut {NoStop}%
\bibitem [{\citenamefont {Vitale}\ \emph {et~al.}(2024)\citenamefont {Vitale},
  \citenamefont {Rath}, \citenamefont {Jurcevic}, \citenamefont {Elben},
  \citenamefont {Branciard},\ and\ \citenamefont {Vermersch}}]{VV24}%
  \BibitemOpen
  \bibfield  {author} {\bibinfo {author} {\bibfnamefont {V.}~\bibnamefont
  {Vitale}}, \bibinfo {author} {\bibfnamefont {A.}~\bibnamefont {Rath}},
  \bibinfo {author} {\bibfnamefont {P.}~\bibnamefont {Jurcevic}}, \bibinfo
  {author} {\bibfnamefont {A.}~\bibnamefont {Elben}}, \bibinfo {author}
  {\bibfnamefont {C.}~\bibnamefont {Branciard}}, \ and\ \bibinfo {author}
  {\bibfnamefont {B.}~\bibnamefont {Vermersch}},\ }\href {\doibase
  10.1103/PRXQuantum.5.030338} {\bibfield  {journal} {\bibinfo  {journal} {PRX
  Quantum}\ }\textbf {\bibinfo {volume} {5}},\ \bibinfo {pages} {030338}
  (\bibinfo {year} {2024})}\BibitemShut {NoStop}%
\bibitem [{\citenamefont {Vermersch}\ \emph {et~al.}(2024)\citenamefont
  {Vermersch}, \citenamefont {Ljubotina}, \citenamefont {Cirac}, \citenamefont
  {Zoller}, \citenamefont {Serbyn},\ and\ \citenamefont {Piroli}}]{BV24}%
  \BibitemOpen
  \bibfield  {author} {\bibinfo {author} {\bibfnamefont {B.}~\bibnamefont
  {Vermersch}}, \bibinfo {author} {\bibfnamefont {M.}~\bibnamefont
  {Ljubotina}}, \bibinfo {author} {\bibfnamefont {J.~I.}\ \bibnamefont
  {Cirac}}, \bibinfo {author} {\bibfnamefont {P.}~\bibnamefont {Zoller}},
  \bibinfo {author} {\bibfnamefont {M.}~\bibnamefont {Serbyn}}, \ and\ \bibinfo
  {author} {\bibfnamefont {L.}~\bibnamefont {Piroli}},\ }\href {\doibase
  10.1103/PhysRevX.14.031035} {\bibfield  {journal} {\bibinfo  {journal} {Phys.
  Rev. X}\ }\textbf {\bibinfo {volume} {14}},\ \bibinfo {pages} {031035}
  (\bibinfo {year} {2024})}\BibitemShut {NoStop}%
\bibitem [{\citenamefont {\.{Z}yczkowski}\ \emph {et~al.}(1998)\citenamefont
  {\.{Z}yczkowski}, \citenamefont {Horodecki}, \citenamefont {Sanpera},\ and\
  \citenamefont {Lewenstein}}]{KZ98}%
  \BibitemOpen
  \bibfield  {author} {\bibinfo {author} {\bibfnamefont {K.}~\bibnamefont
  {\.{Z}yczkowski}}, \bibinfo {author} {\bibfnamefont {P.}~\bibnamefont
  {Horodecki}}, \bibinfo {author} {\bibfnamefont {A.}~\bibnamefont {Sanpera}},
  \ and\ \bibinfo {author} {\bibfnamefont {M.}~\bibnamefont {Lewenstein}},\
  }\href {\doibase 10.1103/PhysRevA.58.883} {\bibfield  {journal} {\bibinfo
  {journal} {Phys. Rev. A}\ }\textbf {\bibinfo {volume} {58}},\ \bibinfo
  {pages} {883} (\bibinfo {year} {1998})}\BibitemShut {NoStop}%
\bibitem [{\citenamefont {Vidal}\ and\ \citenamefont {Tarrach}(1999)}]{GV99}%
  \BibitemOpen
  \bibfield  {author} {\bibinfo {author} {\bibfnamefont {G.}~\bibnamefont
  {Vidal}}\ and\ \bibinfo {author} {\bibfnamefont {R.}~\bibnamefont
  {Tarrach}},\ }\href {\doibase 10.1103/PhysRevA.59.141} {\bibfield  {journal}
  {\bibinfo  {journal} {Phys. Rev. A}\ }\textbf {\bibinfo {volume} {59}},\
  \bibinfo {pages} {141} (\bibinfo {year} {1999})}\BibitemShut {NoStop}%
\bibitem [{\citenamefont {Braunstein}\ \emph {et~al.}(1999)\citenamefont
  {Braunstein}, \citenamefont {Caves}, \citenamefont {Jozsa}, \citenamefont
  {Linden}, \citenamefont {Popescu},\ and\ \citenamefont {Schack}}]{SLB99}%
  \BibitemOpen
  \bibfield  {author} {\bibinfo {author} {\bibfnamefont {S.~L.}\ \bibnamefont
  {Braunstein}}, \bibinfo {author} {\bibfnamefont {C.~M.}\ \bibnamefont
  {Caves}}, \bibinfo {author} {\bibfnamefont {R.}~\bibnamefont {Jozsa}},
  \bibinfo {author} {\bibfnamefont {N.}~\bibnamefont {Linden}}, \bibinfo
  {author} {\bibfnamefont {S.}~\bibnamefont {Popescu}}, \ and\ \bibinfo
  {author} {\bibfnamefont {R.}~\bibnamefont {Schack}},\ }\href {\doibase
  10.1103/PhysRevLett.83.1054} {\bibfield  {journal} {\bibinfo  {journal}
  {Phys. Rev. Lett.}\ }\textbf {\bibinfo {volume} {83}},\ \bibinfo {pages}
  {1054} (\bibinfo {year} {1999})}\BibitemShut {NoStop}%
\bibitem [{\citenamefont {Wen}\ and\ \citenamefont {Kempf}(2023)}]{RYW23}%
  \BibitemOpen
  \bibfield  {author} {\bibinfo {author} {\bibfnamefont {R.~Y.}\ \bibnamefont
  {Wen}}\ and\ \bibinfo {author} {\bibfnamefont {A.}~\bibnamefont {Kempf}},\
  }\href {\doibase 10.1088/1751-8121/ace810} {\bibfield  {journal} {\bibinfo
  {journal} {J. Phys. A: Math. Theor.}\ }\textbf {\bibinfo {volume} {56}},\
  \bibinfo {pages} {335302} (\bibinfo {year} {2023})}\BibitemShut {NoStop}%
\bibitem [{\citenamefont {Parez}\ and\ \citenamefont
  {Witczak-Krempa}(2024)}]{GP24}%
  \BibitemOpen
  \bibfield  {author} {\bibinfo {author} {\bibfnamefont {G.}~\bibnamefont
  {Parez}}\ and\ \bibinfo {author} {\bibfnamefont {W.}~\bibnamefont
  {Witczak-Krempa}},\ }\href {https://arxiv.org/abs/2402.06677} {\enquote
  {\bibinfo {title} {The fate of entanglement},}\ } (\bibinfo {year} {2024}),\
  \bibinfo {note} {arXiv:2402.06677}\BibitemShut {NoStop}%
\bibitem [{\citenamefont {Fine}\ \emph {et~al.}(2005)\citenamefont {Fine},
  \citenamefont {Mintert},\ and\ \citenamefont {Buchleitner}}]{BVF05}%
  \BibitemOpen
  \bibfield  {author} {\bibinfo {author} {\bibfnamefont {B.~V.}\ \bibnamefont
  {Fine}}, \bibinfo {author} {\bibfnamefont {F.}~\bibnamefont {Mintert}}, \
  and\ \bibinfo {author} {\bibfnamefont {A.}~\bibnamefont {Buchleitner}},\
  }\href {\doibase 10.1103/PhysRevB.71.153105} {\bibfield  {journal} {\bibinfo
  {journal} {Phys. Rev. B}\ }\textbf {\bibinfo {volume} {71}},\ \bibinfo
  {pages} {153105} (\bibinfo {year} {2005})}\BibitemShut {NoStop}%
\bibitem [{\citenamefont {Bakshi}\ \emph {et~al.}(2024)\citenamefont {Bakshi},
  \citenamefont {Liu}, \citenamefont {Moitra},\ and\ \citenamefont
  {Tang}}]{AB24}%
  \BibitemOpen
  \bibfield  {author} {\bibinfo {author} {\bibfnamefont {A.}~\bibnamefont
  {Bakshi}}, \bibinfo {author} {\bibfnamefont {A.}~\bibnamefont {Liu}},
  \bibinfo {author} {\bibfnamefont {A.}~\bibnamefont {Moitra}}, \ and\ \bibinfo
  {author} {\bibfnamefont {E.}~\bibnamefont {Tang}},\ }\href
  {https://arxiv.org/abs/2403.16850} {\enquote {\bibinfo {title}
  {High-temperature gibbs states are unentangled and efficiently preparable},}\
  } (\bibinfo {year} {2024}),\ \bibinfo {note} {arXiv:2403.16850}\BibitemShut
  {NoStop}%
\bibitem [{\citenamefont {Ram\'irez}\ \emph {et~al.}(2015)\citenamefont
  {Ram\'irez}, \citenamefont {Rodr\'iguez-Laguna},\ and\ \citenamefont
  {Sierra}}]{GR15}%
  \BibitemOpen
  \bibfield  {author} {\bibinfo {author} {\bibfnamefont {G.}~\bibnamefont
  {Ram\'irez}}, \bibinfo {author} {\bibfnamefont {J.}~\bibnamefont
  {Rodr\'iguez-Laguna}}, \ and\ \bibinfo {author} {\bibfnamefont
  {G.}~\bibnamefont {Sierra}},\ }\href {\doibase
  10.1088/1742-5468/2015/06/P06002} {\bibfield  {journal} {\bibinfo  {journal}
  {J. Stat. Mech.: Theory Exp.}\ }\textbf {\bibinfo {volume} {2015}},\ \bibinfo
  {pages} {P06002} (\bibinfo {year} {2015})}\BibitemShut {NoStop}%
\bibitem [{\citenamefont {Wolf}\ \emph {et~al.}(2008)\citenamefont {Wolf},
  \citenamefont {Verstraete}, \citenamefont {Hastings},\ and\ \citenamefont
  {Cirac}}]{MMW08}%
  \BibitemOpen
  \bibfield  {author} {\bibinfo {author} {\bibfnamefont {M.~M.}\ \bibnamefont
  {Wolf}}, \bibinfo {author} {\bibfnamefont {F.}~\bibnamefont {Verstraete}},
  \bibinfo {author} {\bibfnamefont {M.~B.}\ \bibnamefont {Hastings}}, \ and\
  \bibinfo {author} {\bibfnamefont {J.~I.}\ \bibnamefont {Cirac}},\ }\href
  {\doibase 10.1103/PhysRevLett.100.070502} {\bibfield  {journal} {\bibinfo
  {journal} {Phys. Rev. Lett.}\ }\textbf {\bibinfo {volume} {100}},\ \bibinfo
  {pages} {070502} (\bibinfo {year} {2008})}\BibitemShut {NoStop}%
\bibitem [{\citenamefont {Kim}\ \emph {et~al.}(2024)\citenamefont {Kim},
  \citenamefont {Lavasani},\ and\ \citenamefont {Vijay}}]{YK24}%
  \BibitemOpen
  \bibfield  {author} {\bibinfo {author} {\bibfnamefont {Y.}~\bibnamefont
  {Kim}}, \bibinfo {author} {\bibfnamefont {A.}~\bibnamefont {Lavasani}}, \
  and\ \bibinfo {author} {\bibfnamefont {S.}~\bibnamefont {Vijay}},\ }\href
  {https://arxiv.org/abs/2408.00066} {\enquote {\bibinfo {title} {Persistent
  topological negativity in a high-temperature mixed-state},}\ } (\bibinfo
  {year} {2024}),\ \bibinfo {note} {arXiv:2408.00066}\BibitemShut {NoStop}%
\bibitem [{\citenamefont {Piroli}\ and\ \citenamefont {Cirac}(2020)}]{LP20}%
  \BibitemOpen
  \bibfield  {author} {\bibinfo {author} {\bibfnamefont {L.}~\bibnamefont
  {Piroli}}\ and\ \bibinfo {author} {\bibfnamefont {J.~I.}\ \bibnamefont
  {Cirac}},\ }\href {\doibase 10.1103/PhysRevLett.125.190402} {\bibfield
  {journal} {\bibinfo  {journal} {Phys. Rev. Lett.}\ }\textbf {\bibinfo
  {volume} {125}},\ \bibinfo {pages} {190402} (\bibinfo {year}
  {2020})}\BibitemShut {NoStop}%
\bibitem [{\citenamefont {Pichler}\ \emph {et~al.}(2013)\citenamefont
  {Pichler}, \citenamefont {Schachenmayer}, \citenamefont {Daley},\ and\
  \citenamefont {Zoller}}]{HP13}%
  \BibitemOpen
  \bibfield  {author} {\bibinfo {author} {\bibfnamefont {H.}~\bibnamefont
  {Pichler}}, \bibinfo {author} {\bibfnamefont {J.}~\bibnamefont
  {Schachenmayer}}, \bibinfo {author} {\bibfnamefont {A.~J.}\ \bibnamefont
  {Daley}}, \ and\ \bibinfo {author} {\bibfnamefont {P.}~\bibnamefont
  {Zoller}},\ }\href {\doibase 10.1103/PhysRevA.87.033606} {\bibfield
  {journal} {\bibinfo  {journal} {Phys. Rev. A}\ }\textbf {\bibinfo {volume}
  {87}},\ \bibinfo {pages} {033606} (\bibinfo {year} {2013})}\BibitemShut
  {NoStop}%
\bibitem [{\citenamefont {Skinner}\ \emph {et~al.}(2019)\citenamefont
  {Skinner}, \citenamefont {Ruhman},\ and\ \citenamefont {Nahum}}]{BS19}%
  \BibitemOpen
  \bibfield  {author} {\bibinfo {author} {\bibfnamefont {B.}~\bibnamefont
  {Skinner}}, \bibinfo {author} {\bibfnamefont {J.}~\bibnamefont {Ruhman}}, \
  and\ \bibinfo {author} {\bibfnamefont {A.}~\bibnamefont {Nahum}},\ }\href
  {\doibase 10.1103/PhysRevX.9.031009} {\bibfield  {journal} {\bibinfo
  {journal} {Phys. Rev. X}\ }\textbf {\bibinfo {volume} {9}},\ \bibinfo {pages}
  {031009} (\bibinfo {year} {2019})}\BibitemShut {NoStop}%
\bibitem [{\citenamefont {Fuji}\ and\ \citenamefont {Ashida}(2020)}]{YF20}%
  \BibitemOpen
  \bibfield  {author} {\bibinfo {author} {\bibfnamefont {Y.}~\bibnamefont
  {Fuji}}\ and\ \bibinfo {author} {\bibfnamefont {Y.}~\bibnamefont {Ashida}},\
  }\href {\doibase 10.1103/PhysRevB.102.054302} {\bibfield  {journal} {\bibinfo
   {journal} {Phys. Rev. B}\ }\textbf {\bibinfo {volume} {102}},\ \bibinfo
  {pages} {054302} (\bibinfo {year} {2020})}\BibitemShut {NoStop}%
\bibitem [{\citenamefont {Veitch}\ \emph {et~al.}(2014)\citenamefont {Veitch},
  \citenamefont {Mousavian}, \citenamefont {Gottesman},\ and\ \citenamefont
  {Emerson}}]{VV14}%
  \BibitemOpen
  \bibfield  {author} {\bibinfo {author} {\bibfnamefont {V.}~\bibnamefont
  {Veitch}}, \bibinfo {author} {\bibfnamefont {S.~A.~H.}\ \bibnamefont
  {Mousavian}}, \bibinfo {author} {\bibfnamefont {D.}~\bibnamefont
  {Gottesman}}, \ and\ \bibinfo {author} {\bibfnamefont {J.}~\bibnamefont
  {Emerson}},\ }\href {\doibase 10.1088/1367-2630/16/1/013009} {\bibfield
  {journal} {\bibinfo  {journal} {New J. Phys.}\ }\textbf {\bibinfo {volume}
  {16}},\ \bibinfo {pages} {013009} (\bibinfo {year} {2014})}\BibitemShut
  {NoStop}%
\bibitem [{\citenamefont {Howard}\ \emph {et~al.}(2014)\citenamefont {Howard},
  \citenamefont {Wallman}, \citenamefont {Veitch},\ and\ \citenamefont
  {Emerson}}]{MH14}%
  \BibitemOpen
  \bibfield  {author} {\bibinfo {author} {\bibfnamefont {M.}~\bibnamefont
  {Howard}}, \bibinfo {author} {\bibfnamefont {J.}~\bibnamefont {Wallman}},
  \bibinfo {author} {\bibfnamefont {V.}~\bibnamefont {Veitch}}, \ and\ \bibinfo
  {author} {\bibfnamefont {J.}~\bibnamefont {Emerson}},\ }\href {\doibase
  10.1038/nature13460} {\bibfield  {journal} {\bibinfo  {journal} {Nature}\
  }\textbf {\bibinfo {volume} {510}},\ \bibinfo {pages} {351} (\bibinfo {year}
  {2014})}\BibitemShut {NoStop}%
\bibitem [{\citenamefont {Liu}\ and\ \citenamefont {Winter}(2022)}]{ZWL22}%
  \BibitemOpen
  \bibfield  {author} {\bibinfo {author} {\bibfnamefont {Z.-W.}\ \bibnamefont
  {Liu}}\ and\ \bibinfo {author} {\bibfnamefont {A.}~\bibnamefont {Winter}},\
  }\href {\doibase 10.1103/PRXQuantum.3.020333} {\bibfield  {journal} {\bibinfo
   {journal} {PRX Quantum}\ }\textbf {\bibinfo {volume} {3}},\ \bibinfo {pages}
  {020333} (\bibinfo {year} {2022})}\BibitemShut {NoStop}%
\bibitem [{\citenamefont {Ollivier}\ and\ \citenamefont {Zurek}(2001)}]{HO01}%
  \BibitemOpen
  \bibfield  {author} {\bibinfo {author} {\bibfnamefont {H.}~\bibnamefont
  {Ollivier}}\ and\ \bibinfo {author} {\bibfnamefont {W.~H.}\ \bibnamefont
  {Zurek}},\ }\href {\doibase 10.1103/PhysRevLett.88.017901} {\bibfield
  {journal} {\bibinfo  {journal} {Phys. Rev. Lett.}\ }\textbf {\bibinfo
  {volume} {88}},\ \bibinfo {pages} {017901} (\bibinfo {year}
  {2001})}\BibitemShut {NoStop}%
\bibitem [{\citenamefont {Datta}\ \emph {et~al.}(2008)\citenamefont {Datta},
  \citenamefont {Shaji},\ and\ \citenamefont {Caves}}]{AD08}%
  \BibitemOpen
  \bibfield  {author} {\bibinfo {author} {\bibfnamefont {A.}~\bibnamefont
  {Datta}}, \bibinfo {author} {\bibfnamefont {A.}~\bibnamefont {Shaji}}, \ and\
  \bibinfo {author} {\bibfnamefont {C.~M.}\ \bibnamefont {Caves}},\ }\href
  {\doibase 10.1103/PhysRevLett.100.050502} {\bibfield  {journal} {\bibinfo
  {journal} {Phys. Rev. Lett.}\ }\textbf {\bibinfo {volume} {100}},\ \bibinfo
  {pages} {050502} (\bibinfo {year} {2008})}\BibitemShut {NoStop}%
\bibitem [{\citenamefont {Radhakrishnan}\ \emph {et~al.}(2020)\citenamefont
  {Radhakrishnan}, \citenamefont {Lauri\`ere},\ and\ \citenamefont
  {Byrnes}}]{CR20}%
  \BibitemOpen
  \bibfield  {author} {\bibinfo {author} {\bibfnamefont {C.}~\bibnamefont
  {Radhakrishnan}}, \bibinfo {author} {\bibfnamefont {M.}~\bibnamefont
  {Lauri\`ere}}, \ and\ \bibinfo {author} {\bibfnamefont {T.}~\bibnamefont
  {Byrnes}},\ }\href {\doibase 10.1103/PhysRevLett.124.110401} {\bibfield
  {journal} {\bibinfo  {journal} {Phys. Rev. Lett.}\ }\textbf {\bibinfo
  {volume} {124}},\ \bibinfo {pages} {110401} (\bibinfo {year}
  {2020})}\BibitemShut {NoStop}%
\bibitem [{\citenamefont {Brunner}\ \emph {et~al.}(2014)\citenamefont
  {Brunner}, \citenamefont {Cavalcanti}, \citenamefont {Pironio}, \citenamefont
  {Scarani},\ and\ \citenamefont {Wehner}}]{NB14}%
  \BibitemOpen
  \bibfield  {author} {\bibinfo {author} {\bibfnamefont {N.}~\bibnamefont
  {Brunner}}, \bibinfo {author} {\bibfnamefont {D.}~\bibnamefont {Cavalcanti}},
  \bibinfo {author} {\bibfnamefont {S.}~\bibnamefont {Pironio}}, \bibinfo
  {author} {\bibfnamefont {V.}~\bibnamefont {Scarani}}, \ and\ \bibinfo
  {author} {\bibfnamefont {S.}~\bibnamefont {Wehner}},\ }\href {\doibase
  10.1103/RevModPhys.86.419} {\bibfield  {journal} {\bibinfo  {journal} {Rev.
  Mod. Phys.}\ }\textbf {\bibinfo {volume} {86}},\ \bibinfo {pages} {419}
  (\bibinfo {year} {2014})}\BibitemShut {NoStop}%
\bibitem [{\citenamefont {Tura}\ \emph {et~al.}(2014)\citenamefont {Tura},
  \citenamefont {Augusiak}, \citenamefont {Sainz}, \citenamefont
  {V^^c3^^a9rtesi}, \citenamefont {Lewenstein},\ and\ \citenamefont
  {Ac^^c3^^adn}}]{JT14}%
  \BibitemOpen
  \bibfield  {author} {\bibinfo {author} {\bibfnamefont {J.}~\bibnamefont
  {Tura}}, \bibinfo {author} {\bibfnamefont {R.}~\bibnamefont {Augusiak}},
  \bibinfo {author} {\bibfnamefont {A.~B.}\ \bibnamefont {Sainz}}, \bibinfo
  {author} {\bibfnamefont {T.}~\bibnamefont {V^^c3^^a9rtesi}}, \bibinfo
  {author} {\bibfnamefont {M.}~\bibnamefont {Lewenstein}}, \ and\ \bibinfo
  {author} {\bibfnamefont {A.}~\bibnamefont {Ac^^c3^^adn}},\ }\href {\doibase
  10.1126/science.1247715} {\bibfield  {journal} {\bibinfo  {journal}
  {Science}\ }\textbf {\bibinfo {volume} {344}},\ \bibinfo {pages} {1256}
  (\bibinfo {year} {2014})}\BibitemShut {NoStop}%
\bibitem [{\citenamefont {Schmied}\ \emph {et~al.}(2016)\citenamefont
  {Schmied}, \citenamefont {Bancal}, \citenamefont {Allard}, \citenamefont
  {Fadel}, \citenamefont {Scarani}, \citenamefont {Treutlein},\ and\
  \citenamefont {Sangouard}}]{RS16}%
  \BibitemOpen
  \bibfield  {author} {\bibinfo {author} {\bibfnamefont {R.}~\bibnamefont
  {Schmied}}, \bibinfo {author} {\bibfnamefont {J.-D.}\ \bibnamefont {Bancal}},
  \bibinfo {author} {\bibfnamefont {B.}~\bibnamefont {Allard}}, \bibinfo
  {author} {\bibfnamefont {M.}~\bibnamefont {Fadel}}, \bibinfo {author}
  {\bibfnamefont {V.}~\bibnamefont {Scarani}}, \bibinfo {author} {\bibfnamefont
  {P.}~\bibnamefont {Treutlein}}, \ and\ \bibinfo {author} {\bibfnamefont
  {N.}~\bibnamefont {Sangouard}},\ }\href {\doibase 10.1126/science.aad8665}
  {\bibfield  {journal} {\bibinfo  {journal} {Science}\ }\textbf {\bibinfo
  {volume} {352}},\ \bibinfo {pages} {441} (\bibinfo {year}
  {2016})}\BibitemShut {NoStop}%
\bibitem [{\citenamefont {Chen}\ and\ \citenamefont {Grover}(2024)}]{YHC24}%
  \BibitemOpen
  \bibfield  {author} {\bibinfo {author} {\bibfnamefont {Y.-H.}\ \bibnamefont
  {Chen}}\ and\ \bibinfo {author} {\bibfnamefont {T.}~\bibnamefont {Grover}},\
  }\href {\doibase 10.1103/PhysRevLett.132.170602} {\bibfield  {journal}
  {\bibinfo  {journal} {Phys. Rev. Lett.}\ }\textbf {\bibinfo {volume} {132}},\
  \bibinfo {pages} {170602} (\bibinfo {year} {2024})}\BibitemShut {NoStop}%
\bibitem [{\citenamefont {Horn}\ and\ \citenamefont {Johnson}(2012)}]{RAH12}%
  \BibitemOpen
  \bibfield  {author} {\bibinfo {author} {\bibfnamefont {R.~A.}\ \bibnamefont
  {Horn}}\ and\ \bibinfo {author} {\bibfnamefont {C.~R.}\ \bibnamefont
  {Johnson}},\ }\href@noop {} {\emph {\bibinfo {title} {Matrix Analysis}}}\
  (\bibinfo  {publisher} {Cambridge University Press, Cambridge},\ \bibinfo
  {year} {2012})\BibitemShut {NoStop}%
\end{thebibliography}%

\end{document}